\documentclass[12pt,a4paper,final]{iopart}
\usepackage{iopams}  
\usepackage{graphicx}
\usepackage{hyperref}
\usepackage{cleveref}

\begin{document}
\title[Local Existence]{Local well-posedness of the Einstein-Yang-Mills system in CMCSHGC gauge}
\author{Puskar Mondal}
%\address{Center of Methematical Sciences and Applications, Department of Mathematics, Harvard University}
%\ead{puskar\_mondal@fas.harvard.edu}

\begin{abstract}
We study the local well-posedness of the Einstein-Yang-Mills equations in constant mean extrinsic curvature spatial harmonic generalized Coulomb gauge (CMCSHGC). In this choice of gauge, the complete Einstein-Yang-Mills equations reduce to a coupled elliptic-hyperbolic system. Utilizing the method developed by Andersson and Moncrief \cite{andersson2003elliptic}, we establish the existence of a unique, local, continuous-in-time solution of this coupled system. This yields an `in time' continuation criteria of the solutions which is to be used in the potential future proof of an improved continuation criteria for this coupled system utilizing Moncrief's light cone estimate technique. 
\end{abstract}

\section{Introduction}
Construction of solutions to classical field equations such as the Einstein and Yang-Mills equations is of extreme importance in mathematical physics. This is primarily because the solutions of these equations (or their quantum analogs) describe the three most important fundamental forces present in nature- the gravity, strong and weak forces. If we ignore the quantum aspects for the moment, these equations exhibit interesting dynamics at the pure classical level due to their rich non-linear structure. For pure gravity, this non-linear structure gives rise to the formation of black holes through the concentration of gravitational energy (spacetime curvature). If coupling between the Einstein and Yang-Mills fields is considered, then interesting phenomena occur even in very special geometries solely because of the non-linearities. Striking examples of such phenomena are the existence of soliton-like solutions \cite{bartnik, yau} and non-trivial dynamics \cite{Moncriefrinne} even in spherically symmetric spacetime a very special geometry. Therefore, from a physical perspective, it is extremely important to study the dynamics of these coupled equations. In particular, one is interested in the long-time behavior of these equations and whether the solutions (lying in suitable function spaces) persist for all \textit{time} and preserve the regularity of the initial data. In other words, regular Cauchy data should extend uniquely and continuously to globally defined, singularity-free solutions of the associated field equations on the entire spacetime. If a breakdown of the solutions were to occur that would be pathological in the sense that the underlying theory then turns out to be non-physical. In the context of gravity (or gravity coupled to a source), the question of predictability (or breakdown of such) is closely tied to the \textit{Cosmic Censorship Conjecture} \cite{penrose1999question} of Penrose, which to this day remains open. In the case of gravitational dynamics, it is slightly more subtle due to the fact that singularities do occur due to the collapse of matter and such singularities are in general hidden behind a horizon and therefore cannot be accessed by a time-like observer located in the domain of outer communications. Therefore, the occurrence of a \textit{naked} singularity is considered pathological (or the breakdown of the theory) in the context of gravity. However, before studying the long-time behavior of the solutions of these equations, it is necessary to establish a short-time well-posedness result. In other words, one should be able to prove that local \textit{in time} solutions to these field equations exist and such solutions are unique and depend continuously on the initial data. In doing so one obtains continuation criteria of the solutions and therefore the global existence (or finite time breakdown) depends on satisfying (or not satisfying) this continuation criterion.          

The Einstein and Yang-Mills equations (or its abelian version the Maxwell equations) however contain constraints in addition to evolution equations. Therefore, it is often necessary to rewrite the system, either by extracting a hyperbolic system or performing suitable gauge fixing. The gauge fixing may yield a hyperbolic system such as spacetime harmonic gauge (also called wave gauge or de Donder gauge; the spacetime coordinates $\{x^{\mu}\}$ are forced to satisfy the wave equation $\Box_{\hat{g}}x^{\mu}=0$, $\hat{g}$ being the Lorentzian metric of the spacetime under consideration) which turns the Einstein evolution equations into hyperbolic equations. In particular, Yvonne Choquet Bruhat's \cite{yvone} celebrated result on a local well-posedness theorem for the vacuum Einstein equations was obtained in this gauge. The analog of spacetime harmonic gauge in the context of Yang-Mills theory is the Lorentz gauge, which once again puts the Yang-Mills evolution equations into hyperbolic form \cite{selberg2016null,tesfahun2015local}. In addition to these gauge choices which are naturally adapted to equations in fully spacetime covariant form, there are other gauges suitable for studying the equations on a spacetime $M$ of the product type $\mathbb{R}\times \Sigma$. One of the most convenient ones is the constant mean curvature spatial harmonic gauge (CMCSH gauge) introduced by Andersson and Moncrief \cite{andersson2003elliptic} to study the vacuum Einstein equations. Application of CMCSH gauge turns Einstein's equations into a coupled elliptic-hyperbolic system whose local well-posedness was proven in \cite{andersson2003elliptic}. In the context of Yang-Mills theory, the Coulomb gauge is a natural analog of the spatial harmonic gauge and it was used by Klainerman and Machedon \cite{klainerman1995finite} to study the local and global existence problems of the Yang-Mills equations on Minkowski spacetime. 

In addition to the fundamental studies presented in the previous paragraph, numerous studies in the literature deal with local and global existence problems of the Einstein and Yang-Mills equations. However, we will only describe the studies that are relevant to our problem. Firstly, we mention the work of Lindblad and Rodnianski \cite{lindblad2010global} which used the spacetime harmonic gauge to obtain a global stability result of Minkowski space. Lefloch and Ma \cite{lefloch2016global} used the same gauge to establish a small data global existence result of the Einstein-coupled-to-Klein-Gordon system. In addition to the stability problem of Minkowski space, spacetime harmonic gauge is also used in the context of the cosmological stability problem by \cite{rodnianski2009stability,speck2012nonlinear}. In addition to these examples, there are numerous other studies (e.g., \cite{lindblad2017asymptotic, ettinger2017sharp, ames2017class}) that utilize this particular choice of gauge. Andersson and Moncrief \cite{andersson2011einstein} proved an asymptotic stability result of the Milne universe utilizing the CMCSH gauge. Later Andersson and Fajman used it to prove a small data global existence result for the Einstein-Vlassov system on a Milne spacetime \cite{andersson2020nonlinear}. In addition, \cite{wang2014rough} obtained a rough data local well-posedness result for the vacuum Einstein equations, \cite{fajman2016local} proved a local well-posedness result of Einstein-coupled- with-Vlasov-matter, \cite{fajman2021slowly} studied the stability of the Milne universe in the presence of a perfect fluid, \cite{fajman2021attractors} studied the asymptotic stability of the Milne universe coupled to a Klein-Gordon field, etc. There are numerous other studies where CMCSH gauge is used to study the vacuum gravity problem or gravity coupled to matter fields (e.g., \cite{moncrief2019could, mondal2019attractors,mondal2021linear}). A substantial amount of study has also been done in the Yang-Mills sector. Eardley and Moncrief \cite{eardley1982global, eardley1982global2} established the global existence of Yang-Mills fields on a Minkowski background in temporal gauge. Later, \cite{chrusciel1997global} extended the global existence result to globally hyperbolic curved spacetimes in the same gauge. Using Coulomb gauge, Klainerman and Machedon \cite{klainerman1995finite}, as mentioned earlier, established a global existence result in energy norm. Later Tao \cite{tao2003local} presented a below-energy-norm local existence theorem for the Yang-Mills equations on Minkowski space in temporal gauge. There are countless other studies which are less relevant in the current context. 

This paper was motivated in part by the desire to adapt the Andersson and Moncrief \cite{andersson2003elliptic} argument to including a Yang-Mills source in Einstein's equations. This introduces an extra gauge choice for the Yang-Mills sector namely the \textit{generalized Coulomb} gauge. In addition, we also want to study three particular problems in the near future: the fully non-linear stability of the Milne universe for Yang-Mills fields, the global existence problem for the Yang-Mills equations on a curved spacetime using Moncrief's light cone estimate technique (in prep.), and obtaining an improved breakdown criteria for the coupled Einstein-Yang-mills equations (similar to the ones provided by \cite{klainerman2010breakdown, shao2011breakdown}). However, before proceeding to tackle such problems, it is mandatory to establish a local existence theorem for the fully coupled Einstein-Yang-Mills equations or the Yang-Mills equations on a curved background. However, If we solve the more general problem of fully coupled Einstein-Yang-Mills equations, the latter follows automatically. The local existence theorem automatically provides a continuation criterion that is crucial to establishing a global existence result. As an example, \cite{eardley1982global,eardley1982global2} first proved a local existence theorem for the Yang-Mills equations and obtained a continuation criterion as the non-blow-up of a certain Sobolev norm. Thereafter they proved, using a light cone estimate technique on the Minkowski background, that the same Sobolev norm that appears in the continuation criterion of the local existence theorem can not blow up in finite time thereby yielding global existence. The current article aims to complete the first part of the argument in the case of the coupled Einstein-Yang-Mills equations. There are a few minor difficulties one needs to account for in the fully coupled system in our choice of gauge. One of the minor problems is the \textit{Gribov} degeneracies associated with the spatial harmonic gauge of gravity and generalized Coulomb gauge of Yang-Mills theory \cite{moncrief1979gribov, chodos1980geometrical, fischer1996quantum, singer1978some, gribov2001instability}. This roughly translates to the gauge orbits ceasing to be transversal to the chosen gauge slice. In addition to this problem, we also need to deal with subtle issues associated with the regularity and closing of the energy estimates in the fully coupled system. We will discuss this in relevant chapters. The remaining is a routine procedure to obtain a local well-posedness theorem.  

We start with deriving the gauge fixed coupled Einstein-Yang-Mills equations. The gauge fixing reduces the system to a coupled elliptic-hyperbolic system. The gauge variables (lapse function, shift vector field, and the shifted temporal component of the Yang-Mills connection) are determined as non-local functionals of the dynamical variables by solving the relevant elliptic equations. We then obtain energy inequalities which together with a contraction mapping argument yields the existence of a unique solution. The in-time continuity and Cauchy stability follow in a standard way. The resulting theorem provides continuation criteria of the local solution.   

\section{Notations}
The `$n+1$' dimensional $C^{\infty}$ spacetime manifold is denoted by $\hat{M}$. We are interested in spacetime manifolds $\hat{M}$ with the product topology $\mathbb{R}\times \Sigma$, where $\Sigma$ denotes the $n$-dimensional closed spatial manifold diffeomorphic to a Cauchy hypersurface (i.e., every inextendible causal curve intersects $\Sigma$ exactly once). We denote the space of Riemannian metrics on $\Sigma$ by $\mathcal{M}_{\Sigma}$. We explicitly work in the $L^{2}$ (with respect to a background metric) Sobolev space $H^{s}$ for $s>\frac{n}{2}+1$. We will consider tensors as sections of a rank-$k$ ($k\geq 0$) vector bundle $\mathcal{B}^{k}$ of tensors over $(\Sigma,g)$, where $g\in \mathcal{M}_{\Sigma}$ is the dynamical metric that solves the Einstein's equations written in $n+1$ form. The $L^{2}$ inner product on a fibre of such a rank-$k$ bundle is defined as
\begin{eqnarray}
\langle u|v\rangle_{L^{2}}:=\int_{\Sigma}u_{i_{1}i_{2}i_{3}....i_{k}}v_{j_{1}j_{2}j_{3}....j_{k}}\gamma^{i_{1}j_{1}}\gamma^{i_{1}j_{2}}\gamma^{i_{3}j_{3}}....\gamma^{i_{k}j_{k}}\mu_{g},
\end{eqnarray}
where $\gamma\in \mathcal{M}_{\Sigma}$ is a $C^{\infty}$ background metric lying in a small enough neighbourhood of $g$ in a suitable space (this is possible since $C^{\infty}$ is dense in Sobolev spaces for compact manifolds) and $\mu_{g}=\sqrt{\det(g_{ij})}dx^{1}\wedge dx^{2}\wedge dx^{3}\wedge.........\wedge dx^{n}$ is the volume form associated with the metric $g$. 
The standard norms are defined naturally as follows
\begin{eqnarray}
||u||_{L^{2}}:=\langle u|u\rangle^{1/2}_{L^{2}}\\
||u||_{L^{\infty}}:=\sup_{\Sigma}(u_{i_{1}i_{2}i_{3}....i_{k}}u_{j_{1}j_{2}j_{3}....j_{k}}\gamma^{i_{1}j_{1}}\gamma^{i_{1}j_{2}}\gamma^{i_{3}j_{3}}....\gamma^{i_{k}j_{k}})^{1/2}
\end{eqnarray}
and so on. We adapt the following definition of a \textit{twisted} rough Laplacian acting on sections of $\mathcal{B}^{k}$
\begin{eqnarray}
\Delta^{\gamma}_{g}u:=-\frac{1}{\mu_{g}}\nabla[\gamma]_{i}(g^{ij}\mu_{g}\nabla[\gamma]_{j}u).
\end{eqnarray}
This Laplacian is self-adjoint with respect to the following inner product on derivatives
\begin{eqnarray}
\langle\nabla[\gamma]u|\nabla[\gamma]v\rangle_{L^{2}}:=\int_{\Sigma}g^{ij}\nabla[\gamma]_{i}u_{i_{1}i_{2}....i_{k}}\nabla[\gamma]_{j}v_{j_{1}j_{2}....j_{k}}\gamma^{i_{1}j_{1}}\gamma^{i_{1}j_{2}}....\gamma^{i_{k}j_{k}}\mu_{g}.
\end{eqnarray}
The covariant derivative with respect to the dynamical metric $g$ will be denoted simply by $\nabla$.
We define the ordinary Laplacian of $g$ in the following way such that it has non-negative spectrum i.e., \begin{eqnarray}
\Delta_{g}\equiv-g^{ij}\nabla_{i}\nabla_{j}.
\end{eqnarray}
To formulate the Yang-Mills theory over the spacetime $\mathbb{R}\times \Sigma$, we first choose a compact semi-simple Lie group $G$. If a section of the principle $G-$bundle defined over $\mathbb{R}\times \Sigma$ is chosen and the connection is pulled back to the base manifold, then it yields a $1-$form field on the base which takes values in the Lie algebra $\mathfrak{g}$ of $G$. Let us consider the dimension of the group $G$ to be $dim_{G}$ and since $\mathfrak{g}:=T_{e}G$, it has a natural vector space structure. Assume that the vector space $\mathfrak{g}$ has a basis $\{\chi_{A}\}_{A=1}^{dim_{G}}$ given by a set of $k\times k$ real valued matrices ($k$ being the dimension of the representation $V$ of the Lie algebra $\mathfrak{g}$). The connection $1-$form field is then defined to be 
\begin{eqnarray}
\hat{A}:=\hat{A}^{A}_{\mu}\chi_{A}dx^{\mu}=\hat{A}^{A}_{\mu}(\chi_{A})^{a}_{b}dx^{\mu}=\hat{A}^{a}~_{b\mu}dx^{\mu},~a,b=1,2,3,...,k.
\end{eqnarray}
From now on by the connection 1-form field $\hat{A}_{\mu}$, we will always mean $\hat{A}^{a}~_{b\mu}$. In the current setting $\hat{A}\in \Omega^{1}(\mathbb{R}\times\Sigma;End(V))$, where $End(V)$ denotes the space of endomorphisms of the vector space $V$. The curvature of this connection is defined to be the Yang-Mills field $F\in \Omega^{2}(\mathbb{R}\times\Sigma;End(V))$
\begin{eqnarray}
\hat{F}^{a}~_{b\mu\nu}:=\partial_{\mu}\hat{A}^{a}~_{b\nu}-\partial_{\nu}\hat{A}^{a}~_{b\mu}+[\hat{A},\hat{A}]^{a}~_{b\mu\nu},
\end{eqnarray}
where the bracket is defined on the Lie algebra $\mathfrak{g}$ and given by commutator of matrices under multiplication. The Yang-Mills coupling constant is set to unity. Since $G$ is compact, it admits a positive definite adjoint invariant metric on $\mathfrak{g}$. We choose a basis of $\mathfrak{g}$ such that this adjoint invariant metric takes the Cartesian form $\delta_{AB}$ and work with representations for which the bases satisfy 
\begin{eqnarray}
-\tr(\chi_{A}\chi_{B})=(\chi_{A})^{a}_{b}(\chi_{A})^{b}_{a}=\delta_{AB}.
\end{eqnarray}
This metric allows us to define a gauge invariant (since it is adjoint invariant) inner product on $\mathfrak{g}$. An $L^{2}$ inner product of $A\in sections(\mathcal{B}^{k};End(V))$ and $B\in sections(\mathcal{B}^{k};End(V))$ is defined as follows 
\begin{eqnarray}
\langle A|B\rangle_{L^{2}}:=\int_{\Sigma}A^{a}~_{bi_{1}i_{2}i_{3}....i_{k}}B^{b}~_{aj_{1}j_{2}j_{3}....j_{k}}\gamma^{i_{1}j_{1}}\gamma^{i_{1}j_{2}}\gamma^{i_{3}j_{3}}....\gamma^{i_{k}j_{k}}\mu_{g},
\end{eqnarray}
where the contraction on the gauge indices (or the internal indices) is performed by means of the adjoint invariant metric on $\mathfrak{g}$. Notice that this inner product is not gauge invariant unless $A$ and $B$ both transform as tensors under gauge transformations.

Under a gauge transformation $\mathcal{G}$, the $\mathfrak{g}$ valued 1-form field $\hat{A}$ transforms as 
\begin{eqnarray}
\hat{A}_{\mu}\mapsto \mathcal{G}\hat{A}_{\mu}\mathcal{G}+\mathcal{G}\partial_{\mu}\mathcal{G}^{-1} 
\end{eqnarray}
and therefore $\hat{A}_{\mu}$ is not a tensor in the sense that it is \textit{not} a $(1,1)$ section of the associated $V-$bundle over $\mathbb{R}\times \Sigma$. For any $\mathfrak{g}$ valued section $\mathcal{K}^{a}~_{b\mu_{1}\mu_{2}\mu_{3}....\mu_{k}}$ of a vector bundle over $\mathbb{R}\times \Sigma$ that transform as a tensor under the gauge transformation, the gauge covariant derivative is defined to be  
\begin{eqnarray}
\mathcal{D}_{\alpha}\mathcal{K}^{a}~_{b\mu_{1}\mu_{2}\mu_{3}....\mu_{k}}\\\nonumber 
:=\nabla_{\alpha}\mathcal{K}^{a}~_{b\mu_{1}\mu_{2}\mu_{3}....\mu_{k}}+A^{a}~_{c\alpha}\mathcal{K}^{c}~_{b\mu_{1}\mu_{2}\mu_{3}....\mu_{k}}-A^{c}~_{b\alpha}\mathcal{K}^{a}~_{c\mu_{1}\mu_{2}\mu_{3}....\mu_{k}}\\\nonumber 
=\nabla_{\alpha}\mathcal{K}^{a}~_{b\mu_{1}\mu_{2}\mu_{3}....\mu_{k}}+[A,\mathcal{K}]^{a}~_{b\mu_{1}\mu_{2}\mu_{3}....\mu_{k}},
\end{eqnarray}
where $\nabla_{\alpha}$ is the ordinary spacetime covariant derivative with respect to a Lorentzian metric on $\mathbb{R}\times \Sigma$. Naturally this notion of gauge covariant derivative passes down to the spatial manifold $\Sigma$.\\     
Lastly, for functions $f(t)\in \mathbb{R}_{\geq 0}$ and $g(t)\in \mathbb{R}_{\geq 0}$ $f(t)\lesssim g(t)$ means $f(t)\leq Cg(t)$ and $f(t)\gtrsim g(t)$ means $f(t)\geq Cg(t)$, and $f(t)\approx g(t)$ implies $C_{1}g(t)\leq f(t)\leq C_{2}g(t)$ for $0<C_{1},C_{2},C<\infty$. The involved constants may depend only on the background geometry. The spaces of symmetric covariant 2-tensors and vector fields on $\Sigma$ are denoted by $S^{0}_{2}(\Sigma)$ and $\mathfrak{X}(\Sigma)$, respectively.

\section{Gauge Fixed Einstein-Yang Mills field equations}
We utilize the ADM formalism that splits the spacetime described by an `$n+1$' dimensional $C^{\infty}$ Lorentzian manifold $\hat{M}$ into $\mathbb{R}\times \Sigma$. As mentioned previously, each level set $\{t\}\times \Sigma$ of the time function $t$ is an orientable $n-$manifold diffeomorphic to a Cauchy hypersurface (assuming the spacetime to be globally hyperbolic) and equipped with a Riemannian metric. Such a split may be executed by introducing a lapse function $N$ and shift vector field $X$ tangent to $\Sigma$ and defined such that
\begin{eqnarray}
\partial_{t}&=&N\mathbf{n}+X,
\end{eqnarray}
where $t$ and $\mathbf{n}$ are time and a hypersurface orthogonal future directed timelike unit vector i.e., $\hat{g}(\mathbf{n},\mathbf{n})=-1$, respectively. By means of the natural projection operator $\Pi:=\hat{g}+\mathbf{n}\otimes \mathbf{n}$, one may decompose every tensor field into their $\Sigma-$parallel and $\Sigma-$orthogonal components. This splitting writes the spacetime metric $\hat{g}$ in local coordinates $\{x^{\alpha}\}_{\alpha=0}^{n}=\{t,x^{1},x^{2},....,x^{n}\}$ as 
\begin{eqnarray}
\label{eq:metric}
\hat{g}&=&-N^{2}dt\otimes dt+g_{ij}(dx^{i}+X^{i}dt)\otimes(dx^{j}+X^{j}dt)
\end{eqnarray} 
and the stress-energy tensor as
\begin{eqnarray}
\mathbf{T}&=&E\mathbf{n}\otimes\mathbf{n}+2\mathbf{J}\odot\mathbf{n}+\mathbf{S},
\end{eqnarray}
where $\mathbf{J}\in \mathfrak{X}(\Sigma)$, $\mathbf{S}\in S^{2}_{0}(\Sigma)$, and $A\odot B=\frac{1}{2}(A\otimes B+B\otimes A$). Here, $\mathfrak{X}(\Sigma)$ and $S^{2}_{0}(\Sigma)$ are the space of vector fields and the space of symmetric covariant 2-tensors, respectively. Here $E:=\mathbf{T}(\mathbf{n},\mathbf{n})$ is the energy density observed by a time-like observer with $n+1$-velocity $\mathbf{n}$, $\mathbf{J}_{i}=-\mathbf{T}(\partial_{i},\mathbf{n})$ is the momentum density, $\mathbf{J}_{i}=-\mathbf{T}(\mathbf{n},\partial_{i})=$ is the energy flux density, and $\mathbf{S}_{ij}=\mathbf{T}(\partial_{i},\partial_{j})$ is the momentum flux density (with respect to the chosen constant $t$ hypersurface $\Sigma$). Both the momentum density and the energy flux density have the same value due to the symmetry of the stress energy tensor. The choice of a spatial slice in the spacetime leads to consideration of the second fundamental form $k_{ij}$ which describes the extrinsic geometry of the slice while embedded in the ambient spacetime. The trace of the second fundamental form ($\tau:=\tr_{g}k$) is the mean extrinsic curvature of the slice, which will play an important role in the analysis. Under such decomposition, the Einstein equations 
\begin{eqnarray}
R_{\mu\nu}-\frac{1}{2}R\hat{g}_{\mu\nu}&=&T_{\mu\nu}
\end{eqnarray}
take the form ($8\pi G=c=1$)
\begin{eqnarray}
\partial_{t}g_{ij}&=&-2Nk_{ij}+L_{X}g_{ij},\\\nonumber
\partial_{t}k_{ij}&=&-\nabla_{i}\nabla_{j}N+N\{R_{ij}+\tr_{g}k k_{ij}-2k_{ik}k^{k}_{j}\\\nonumber
&&-\frac{1}{n-1}(E-S)g_{ij}-\mathbf{S}_{ij}\}+L_{X}k_{ij}
\end{eqnarray}
along with the constraints (Gauss and Codazzi equations)
\begin{eqnarray}
\label{eq:HC}
R(g)-|k|^{2}+(\tr_{g}k)^{2}&=&2E,\\
\label{eq:MC}
\nabla_{j}k^{j}_{i}-\nabla_{i}\tr_{g}k&=&-\mathbf{J}_{i},
\end{eqnarray} 
where $S=g^{ij}\mathbf{S}_{ij}$. The vanishing of the covariant divergence of the stress energy tensor i.e., $\nabla_{\nu}T^{\mu\nu}=0$ is equivalent to the continuity equation and equations of motion of the source
\begin{eqnarray}
\frac{\partial E}{\partial t}&=&L_{X}E+NE\tau-N\nabla_{i}\mathbf{J}^{i}-2\mathbf{J}^{i}\nabla_{i}N+N\mathbf{S}^{ij}k_{ij},\\
\frac{\partial \mathbf{J}^{i}}{\partial t}&=&L_{X}\mathbf{J}^{i}+N\tau \mathbf{J}^{i}-\nabla_{j}(N\mathbf{S}^{ij})+2Nk^{i}_{j}\mathbf{J}^{j}-E\nabla^{i}N.
\end{eqnarray}
We want to study the Einstein-Yang-Mills system. One may for convenience decompose the $n+1$ $\mathfrak{g}$-valued connection $1-$form field $A^{a}~_{b\mu}$ into its component parallel and perpendicular to a constant $t$ hypersurface $\Sigma$ as follows 
\begin{eqnarray}
\hat{A}^{a}~_{b}=A^{a}~_{b}-\hat{g}(A^{a}~_{b},\mathbf{n})\mathbf{n}
\end{eqnarray}
where $A^{a}~_{b}$ is a $\mathfrak{g}$ valued $1-$form field parallel to the spatial manifold $\Sigma$ i.e., $A\in \Omega^{1}(\Sigma;End(V))$. Importantly note that $\hat{A}^{a}~_{bi}=A^{a}~_{bi}$ but $\hat{A}^{a}~_{b}^{i}\neq A^{a}~_{b}^{i}$ unless the shift vector field $X$ vanishes. Similarly, we decompose the Yang-Mills field strength $\hat{F}$ as follows 
\begin{eqnarray}
\hat{F}^{a}~_{b}=\mathcal{E}^{a}~_{b}\otimes \mathbf{n}-\mathbf{n}\otimes\mathcal{E}^{a}~_{b}+F^{a}~_{b},
\end{eqnarray}
where $\mathcal{E}\in \Omega^{1}(\Sigma;End(V))$ is the electric field and $F\in \Omega^{2}(\Sigma;End(V))$ is related to the magnetic field. Once again $\hat{F}^{a}~_{bij}=F^{a}~_{bij}$ but $\hat{F}^{a}~_{b}~^{ij}\neq F^{a}~_{b}~^{ij}$. The stress-energy tensor of an $n+1-$Yang-Mills field $\hat{F}$ is given by
\begin{eqnarray}
T_{\mu\nu}=F^{a}~_{b\mu\alpha}F^{b}~_{a\nu}~^{\alpha}-\frac{1}{4}(F^{a}~_{b\alpha\beta}F^{b}~_{a}~^{\alpha\beta})\hat{g}_{\mu\nu}.
\end{eqnarray}
After projecting onto the spatial hypersurface $\Sigma$, several components of the stress-energy tensor may be evaluated as follows 
\begin{eqnarray}
E=\frac{1}{2}g^{ij}\mathcal{E}^{a}~_{bi} \mathcal{E}^{b}~_{aj}+\frac{1}{4}g^{ik}g^{jl}F^{a}~_{bij}F^{b}~_{akl},~\mathbf{J}_{i}=-g^{jk}F^{a}~_{bij}\mathcal{E}^{b}~_{ak},\\\nonumber 
\mathbf{S}_{ij}=-\mathcal{E}^{a}~_{bi} \mathcal{E}^{b}~_{aj}+\left(\frac{1}{2}g^{kl}\mathcal{E}^{a}~_{bk} \mathcal{E}^{b}~_{al}-\frac{1}{4}g^{km}g^{ln}F^{a}~_{bkl} F^{b}~_{amn}\right)g_{ij}\\\nonumber 
+g^{kl}F^{a}~_{bik} F^{b}~_{ajl}
\end{eqnarray}
Using this available information, the complete system of evolution and constraint equations are expressible as follows 
\begin{eqnarray}
\label{eq:YM1}
\partial_{t}A^{a}~_{bi}=N\mathcal{E}^{a}~_{bi}-F^{a}~_{bij}X^{j}+\partial_{i}A^{a}~_{b0}-[A_{0},A_{i}]^{a}~_{b},\\
\partial_{t}\mathcal{E}^{a}~_{bi}=N\tr_{g}k\mathcal{E}^{a}~_{bi}-(\nabla_{j}N)F^{a}~_{bi}~^{j}-N\nabla_{j}F^{a}~_{bi}~^{j}+N[F_{i}~^{j},A_{j}]^{a}~_{b}\\\nonumber+[\mathcal{E}_{i},A_{0}-A_{k}X^{k}]^{a}~_{b}
+X^{k}\nabla_{k}\mathcal{E}^{a}~_{bi}+\mathcal{E}^{a}~_{bk}\nabla_{i}X^{k}-2Nk^{j}_{i}\mathcal{E}^{a}~_{bj},\\
\partial_{t}g_{ij}=-2Nk_{ij}+L_{X}g_{ij}\\
\partial_{t}k_{ij}=-\nabla_{i}\nabla_{j}N+N\{R_{ij}+\tr_{g}k k_{ij}-2k_{ik}k^{k}_{j}\nonumber+\mathcal{E}^{a}~_{bi} \mathcal{E}^{b}~_{aj}-g^{kl}F^{a}~_{bik}F^{b}~_{akl}\\\nonumber 
-\frac{g^{kl}\mathcal{E}^{a}~_{bk}\mathcal{E}^{b}~_{al}}{n-1}g_{ij}+\frac{g^{km}g^{ln}F^{a}~_{bkl}F^{b}~_{amn}}{2(n-1)}g_{ij}\}+L_{X}k_{ij},\\
\label{eq:HC}
R(g)-|k|^{2}+(\tr_{g}k)^{2}=g^{ij}\mathcal{E}^{a}~_{bi}\mathcal{E}^{b}~_{aj}+\frac{1}{2}g^{ik}g^{jl}F^{a}~_{bij}F^{b}~_{akl},\\
\label{eq:MC}
\nabla_{j}k^{j}_{i}-\nabla_{i}\tr_{g}k=g^{jk}F^{a}~_{bij}\mathcal{E}^{b}~_{ak},\\
\label{eq:GLCT}
g^{ij}\nabla_{i}\mathcal{E}^{a}~_{bj}+g^{ij}[A_{i},\mathcal{E}_{j}]^{a}~_{b}=0.
\end{eqnarray}
This system is not hyperbolic since the Ricci tensor of the spatial metric and the divergence of the spatial Yang-Mills curvature contain terms that obstruct the ellipticity of the maps $g\mapsto $Ricci$[g]$ and $A\mapsto \nabla_{j}F^{a}~_{bi}~^{j}$. Therefore we need to extract a hyperbolic system of evolution equations by fixing a suitable gauge, that is, remove these obstructions. For the gravitational sector, we use the constant mean extrinsic curvature spatial harmonic gauge. Setting the mean extrinsic curvature ($\tau$) of the hypersurface $M$ to a time-dependent constant allows it to be a suitable time function. This choice of temporal gauge yields an elliptic equation for the lapse function $N$. However, not all spacetimes admit a constant mean extrinsic curvature (CMC) hypersurface. The existence of a CMC slice is far from obvious and is a part of active mathematical research (for details see \cite{bartnik1988remarks, galloway2018existence, rendall1996constant, andersson1999existence}). This gauge condition reads 
\begin{eqnarray}
\tau:=\tr_{g}k=monotonic~function~of~time~alone.
\end{eqnarray}
This choice allows $\tau$ to play the role of time i.e.,
\begin{eqnarray}
t=monotonic~function~of~\tau.
\end{eqnarray}
To utilize the property $\partial_{i}\tau=0$, one may simply compute the entity $\frac{\partial\tau}{\partial t}$ to yield
\begin{eqnarray}
\partial_{t}\tau\\\nonumber 
=-g^{ij}\nabla_{i}\nabla_{j}N+\left(|k|^{2}_{g}+\frac{n-2}{n-1}g^{kl}\mathcal{E}^{a}~_{bk}\mathcal{E}^{b}~_{al}+\frac{1}{2(n-1)}g^{kl}g^{mn}F^{a}~_{bkm}F^{b}~_{aln}\right)N
\end{eqnarray}
which is nothing but an elliptic equation for the lapse function $N$. 
Once we have fixed the temporal gauge, we need to fix the spatial gauge which would yield an equation for the shift vector field. Once again, we follow the work of \cite{andersson2004future, andersson2011einstein} regarding spatial gauge fixing. Let $\zeta: (M,g)\to (M,\gamma)$ be a harmonic map with the Dirichlet energy $\frac{1}{2}\int_{M}g^{ij}\frac{\partial \zeta^{k}}{\partial x^{i}}\frac{\partial \zeta^{l}}{\partial x^{j}}\gamma_{kl}\mu_{g}$, where $~\gamma$ is a fixed background Riemannian metric. Since the harmonic maps are the critical points of the Dirichlet energy functional, $\zeta$ satisfies the following formal Euler-Lagrange equation 
\begin{eqnarray}
g^{ij}\left(\partial_{i}\partial_{j}\zeta^{k}-\Gamma[g]_{ij}^{l}\partial_{l} \zeta^{k}+\Gamma[\gamma]_{\alpha\beta}^{k}\partial_{i} \zeta^{\alpha}\partial_{j} \zeta^{\beta}\right)=0.
\end{eqnarray}
Now, we fix the gauge by imposing the condition that $\zeta=id$, which leads to the following equation 
\begin{eqnarray}
\label{eq:sh}
-g^{ij}(\Gamma[g]_{ij}^{k}-\Gamma[\gamma]_{ij}^{k})&=&0.
\end{eqnarray}
where $\Gamma[\gamma]_{ij}^{k}$ is the connection with respect to a background Riemannian metric $\gamma$. Choice of this spatial harmonic gauge yields the following elliptic equation for the shift vector field $X$ after time differentiating equation (\ref{eq:sh})  
\begin{eqnarray}
\Delta_{g}X^{i}-R^{i}_{j}X^{j}&=&-2k^{ij}\nabla_{j}N-2Ng^{ik}g^{lj}F^{a}~_{bkl} \mathcal{E}^{b}~_{aj}+\tau\nabla^{i}N\\\nonumber
&&+(2Nk^{jk}-2\nabla^{j}X^{k})(\Gamma[g]^{i}_{jk}-\Gamma[\gamma]^{i}_{jk}).
\end{eqnarray} 
The Ricci$[g]$ is expressible as follows 
\begin{eqnarray}
R[g]_{ij}=\frac{1}{2}\Delta^{\gamma}_{g}g_{ij}+\alpha_{ij}+\mathcal{N}[g,\partial g]_{ij},
\end{eqnarray}
where
\begin{eqnarray}
\label{eq:riccinonlinear}
\mathcal{N}_{ij}=\frac{1}{2}\left(g_{il}g^{mn}R[\gamma]^{l}_{mjn}+g_{jl}g^{mn}R[\gamma]^{l}_{min}\right)+\frac{1}{2}g^{mn}\gamma^{ls}\left(\nabla[\gamma]_{j}g_{ns}\nabla[\gamma]_{l}g_{im}\right.\\\nonumber 
\left.+\nabla[\gamma]_{i}g_{lm}\nabla[\gamma]_{s}g_{jn}-\frac{1}{2}\nabla[\gamma]_{j}g_{ns}\nabla[\gamma]_{i}g_{lm}+\nabla[\gamma]_{m}g_{il}\nabla[\gamma]_{n}g_{js}\right.\\\nonumber 
\left.-\nabla[\gamma]_{m}g_{il}\nabla[\gamma]_{s}g_{jn}\right)-\frac{1}{2}V^{l}\nabla[\gamma]_{l}g_{ij}
\end{eqnarray}
and 
\begin{eqnarray}
\alpha_{ij}:=\frac{1}{2}(\nabla_{i}V_{j}+\nabla_{j}V_{i}),~V^{k}:=g^{ij}(\Gamma[g]_{ij}^{k}-\Gamma[\gamma]_{ij}^{k}).
\end{eqnarray}
In the SH gauge therefore the term in $\alpha_{ij}$ obstructing the ellipticity of the map $g\mapsto$ Ricci$[g]$ vanishes yielding a hyperbolic system. 

The remaining gauge choice is the gauge fixing of the Yang-Mills sector. Before moving to the gauge fixing, we need to mention a small subtlety that arises in the spatial harmonic gauge. Notice that in this gauge the shift vector field $X\neq 0$ in general and therefore, the evolution equation of $A^{a}~_{bi}$ (\ref{eq:YM1}) includes a term $F^{a}~_{bij}X^{j}$ which contains the first derivative of $A$. This term is therefore at the level of principal terms concerning regularity. While writing down the energy expression, one would ideally want this term to be canceled point-wise. However, that does not happen in our case. Therefore, this term potentially obstructs the closing of the energy argument. However, this involves a gauge variable $X$ and one expects that this term should not create an additional problem. After carefully expressing the term $F^{a}~_{bij}X^{j}$ we observe that one may obtain the following evolution equation for the vector potential $A$
\begin{eqnarray}
\partial_{t}A^{a}~_{bi}&=&N\mathcal{E}^{a}~_{bi}+\partial_{i}(A^{a}~_{b0}-A^{a}~_{bj}X^{j})+[A_{i},A_{0}-A_{j}X^{j}]^{a}~_{b}\\\nonumber 
&&+X^{j}\nabla[\gamma]_{j}A^{a}~_{bi}\nonumber+A^{a}~_{bj}\nabla[\gamma]_{i}X^{j}
\end{eqnarray}
Now instead of treating the temporal component $A^{a}_{b0}$ as the Yang-Mills gauge variable, it suffices to treat $\varphi^{a}_{b}:=A^{a}~_{b0}-A^{a}~_{bj}X^{j}$ as the new gauge variable. By fixing a suitable gauge, we will obtain a necessary elliptic equation for $\varphi^{a}_{b}$. This issue of regularity will be very crucial while studying the elliptic equation.
We will use a version of the ordinary Coulomb gauge that is used in Maxwell theory (or sometimes in Yang-Mills theory). The ordinary Coulomb gauge condition simply states the vanishing of the divergence of the vector potential $A^{a}~_{bi}$. However, we notice that $A^{a}~_{bi}$ does not transform as a tensor. Therefore, we define the \textit{generalized} Coulomb gauge condition in the following way 
\begin{eqnarray}
\label{eq:GC}
g^{ij}\nabla^{\hat{A}}[\gamma]_{j}(A_{i}-\hat{A}_{i}):=g^{ij}\nabla[\gamma]_{j}(A_{i}-\hat{A}_{i})+g^{ij}[\hat{A}_{j},(A_{i}-\hat{A}_{i})]=0,
\end{eqnarray}
where $\hat{A}$ is a $C^{\infty}$ background connection and $\nabla^{\hat{A}}[\gamma]$ is the gauge covariant derivative with respect to the background connection $\hat{A}$ and the background metric $\gamma$. The time derivative of this gauge condition yields an elliptic equation for $\varphi^{a}_{b}$. Explicit calculations yield
\begin{eqnarray}
g^{ij}\nabla[\gamma]^{A}_{i}\nabla[\gamma]^{A}_{j}\varphi^{a}~_{b}+g^{ij}[\hat{A}_{i}-A_{i},\nabla^{A}_{j}\varphi]^{a}~_{b}+\nonumber
g^{ij}\nabla_{i}N\mathcal{E}^{a}~_{bj}-Ng^{ij}[A_{i},\mathcal{E}_{j}]^{a}~_{b}\\\nonumber+g^{ij}[\hat{A}_{i},N\mathcal{E}_{j}]^{a}~_{b}
+g^{ij}(\Gamma[g]^{k}_{ij}-\Gamma[\gamma]^{k}_{ij})N\mathcal{E}^{a}~_{bk}-X^{k}\nabla[\gamma]_{k}g^{ij}\nabla[\gamma]_{i}(A^{a}~_{bj}\\\nonumber-\hat{A}^{a}~_{bj})-X^{k}\nabla[\gamma]_{k}g^{ij}[\hat{A}_{i},A_{j}-\hat{A}_{j}]^{a}~_{b} +g^{ij}\nabla[\gamma]_{i}X^{k}\nabla[\gamma]_{k}A^{a}~_{bj}\\\nonumber+g^{ij}\nabla[\gamma]_{i}A^{a}~_{bk}\nabla[\gamma]_{j}X^{k}+g^{ij}A^{a}~_{bk}\nabla[\gamma]_{i}\nabla[\gamma]_{j}X^{k}-g^{ij}R[\gamma]^{l}_{jik}A^{a}~_{bl}X^{k} \\\nonumber+g^{ij}[\hat{A}_{i},A_{k}]^{a}~_{b}\nabla[\gamma]_{j}X^{k}+(2Nk^{ij}-\nabla^{i}X^{j}-\nabla^{j}X^{i})\nabla[\gamma]^{\hat{A}}_{i}(A^{a}~_{bj}-\hat{A}^{a}~_{bj})=0.
\end{eqnarray}
Note that this equation is consistent in terms of regularity. If we assume that $(g,k,A,\mathcal{E})\in H^{s}\times H^{s-1}\times H^{s}\times H^{s-1},~s>\frac{n}{2}+1$, then by the elliptic equations at hand, we automatically obtain $(N,X,\varphi)\in H^{s+1}\times H^{s+1}\times H^{s+1}$. However, if we had used $A_{0}$ instead of $\varphi:=A_{0}-A_{i}X^{i}$, then the time derivative of the generalized Coulomb gauge condition would yield an elliptic equation for $A_{0}$ which would contain the second derivative of $A$ as a source term requiring $A_{0}$ to be in $H^{s+2}$ and therefore the energy estimates for the evolution equations (where a derivative of $A_{0}$ is present) would not close. In a sense, construction of the \textit{shifted} potential $\varphi$ yields a cancellation which removes the problematic term. The term $\nabla_{j}F^{a}~_{bi}~^{j}$ appearing on the right hand side of the evolution equation for the electric field is not generally elliptic as a map $A\mapsto \nabla_{j}F^{a}~_{bi}~^{j}$ since it can be written as follows 
\begin{eqnarray}
\nabla_{j}F^{a}~_{bi}~^{j}=\Delta^{\gamma}_{g}A^{a}~_{bi}+\nabla_{i}\mathcal{C}^{a}~_{b}+\mathcal{SL}^{a}~_{bi},
\end{eqnarray}
where 
\begin{eqnarray}
\mathcal{C}^{a}~_{b}&:=&g^{ij}\nabla^{\hat{A}}[\gamma]_{j}(A^{a}~_{bi}-\hat{A}^{a}~_{bi})\\\nonumber 
&=&g^{ij}\nabla[\gamma]_{j}(A^{a}~_{bi}-\hat{A}^{a}~_{bi})+g^{ij}[\hat{A}_{j},(A_{i}-\hat{A}_{i})]^{a}~_{b}
\end{eqnarray}
and $\mathcal{SL}^{a}_{bi}$ is semi-linear in $A$ and $g$ and expressed as follows 
\begin{eqnarray}
\label{eq:YMNL}
\mathcal{SL}^{a}~_{b}~^{i}\\\nonumber
:=g^{jl}\nabla[\gamma]_{j}g^{ik}\nabla[\gamma]_{i}A^{a}~_{bl}+g^{ik}\nabla_{j}g^{jl}\nabla[\gamma]_{k}A^{a}~_{bl}\nonumber-g^{jl}\nabla[\gamma]_{j}g^{ik}\nabla[\gamma]_{l}A^{a}~_{bk}\\\nonumber -g^{ik}\nabla[\gamma]_{j}g^{jl}\nabla[\gamma]_{k}A^{a}~_{bl} 
+g^{jl}\nabla[\gamma]_{j}g^{ik}[A_{k},A_{l}]^{a}~_{b}+g^{ik}\nabla[\gamma]_{j}g^{jl}[A_{k},A_{l}]^{a}~_{b}\\\nonumber+g^{ik}g^{jl}[\nabla[\gamma]_{j}A_{k},A_{l}]^{a}~_{b}+g^{ik}g^{jl}[A_{k},\nabla[\gamma]_{j}A_{l}]^{a}~_{b}+g^{ik}\nabla[\gamma]_{k}(g^{jl}\nabla[\gamma]_{j}\hat{A}^{a}~_{bl})\\\nonumber 
-g^{ik}\nabla[\gamma]_{k}\left(g^{jl}[\hat{A}_{j},A_{l}-\hat{A}_{l}]^{a}~_{b}\right)+g^{ik}g^{jl}R[\gamma]^{m}~_{ljk}A^{a}~_{bm}.
\end{eqnarray}
Therefore the map $A\mapsto \nabla_{j}F^{a}~_{bi}~^{j}-\nabla_{i}\mathcal{C}^{a}~_{b}$ is elliptic thereby turning the Yang-Mills evolution equations into hyperbolic form. 
In order to construct solutions to the Einstein-Yang-Mills coupled evolution and constraint equations in the chosen gauge, we want to solve the following modified system of evolution equations 
\begin{eqnarray}
\label{eq:evolution1}
\partial_{t}A^{a}~_{bi}=N\mathcal{E}^{a}~_{bi}+\partial_{i}\varphi^{a}~_{b}+[A_{i},\varphi]^{a}~_{b}+X^{j}\nabla[\gamma]_{j}A^{a}~_{bi}\nonumber+A^{a}~_{bj}\nabla[\gamma]_{i}X^{j},\\
\partial_{t}\mathcal{E}^{a}~_{bi}=N\tr_{g}k\mathcal{E}^{a}~_{bi}-(\nabla_{j}N)F^{a}~_{bi}~^{j}-\underbrace{N(\nabla_{j}F^{a}~_{bi}~^{j}-\nabla_{i}\mathcal{C}^{a}~_{b})}_{elliptic~as~an~image~of~a~map~of~A}\\\nonumber+N[F_{i}~^{j},A_{j}]^{a}~_{b}+[\mathcal{E}_{i},\varphi]^{a}~_{b}
+X^{k}\nabla_{k}\mathcal{E}^{a}~_{bi}+\mathcal{E}^{a}~_{bk}\nabla_{i}X^{k}-2Nk^{j}_{i}\mathcal{E}^{a}~_{bj},\\
\partial_{t}g_{ij}=-2Nk_{ij}+L_{X}g_{ij}\\
\label{eq:evolution4}
\partial_{t}k_{ij}=-\nabla_{i}\nabla_{j}N+N\{\underbrace{R_{ij}-\alpha_{ij}}_{elliptic~as~an~image~of~a~map~of~g}+\tr_{g}k k_{ij}-2k_{ik}k^{k}_{j}\\\nonumber+\mathcal{E}^{a}~_{bi} \mathcal{E}^{b}~_{aj}-g^{kl}F^{a}~_{bik}F^{b}~_{akl} 
-\frac{g^{kl}\mathcal{E}^{a}~_{bk}\mathcal{E}^{b}~_{al}}{n-1}g_{ij}+\frac{g^{km}g^{ln}F^{a}~_{bkl}F^{b}~_{amn}}{2(n-1)}g_{ij}\}+L_{X}k_{ij}
\end{eqnarray}
coupled with the elliptic equations for $N,X$, and $\varphi$ needed to preserve the imposed gauge conditions 
\begin{eqnarray}
\label{eq:elliptic}
-g^{ij}\nabla_{i}\nabla_{j}N\nonumber+\left(|k|^{2}_{g}+\frac{n-2}{n-1}g^{kl}\mathcal{E}^{a}~_{bk}\mathcal{E}^{b}~_{al}+\frac{1}{2(n-1)}g^{kl}g^{mn}F^{a}~_{bkm}F^{b}~_{aln}\right)\\
N=1,
\end{eqnarray}
\begin{eqnarray}
\label{eq:elliptic2}
\Delta_{g}X^{i}-R^{i}_{j}X^{j}+L_{X}V^{i}=-2k^{ij}\nabla_{j}N-2Ng^{ik}g^{lj}F^{a}~_{bkl} \mathcal{E}^{b}~_{aj}+\tau\nabla^{i}N\\\nonumber
+(2Nk^{jk}-2\nabla^{j}X^{k})(\Gamma[g]^{i}_{jk}-\Gamma[\gamma]^{i}_{jk}),
\end{eqnarray}
\begin{eqnarray}
\label{eq:elliptic3}
g^{ij}\nabla[\gamma]^{A}_{i}\nabla[\gamma]^{A}_{j}\varphi^{a}~_{b}+g^{ij}[\hat{A}_{i}-A_{i},\nabla^{A}_{j}\varphi]^{a}~_{b}+\nonumber
g^{ij}\nabla_{i}N\mathcal{E}^{a}~_{bj}-Ng^{ij}[A_{i},\mathcal{E}_{j}]^{a}~_{b}\\+g^{ij}[\hat{A}_{i},N\mathcal{E}_{j}]^{a}~_{b}
+g^{ij}(\Gamma[g]^{k}_{ij}-\Gamma[\gamma]^{k}_{ij})N\mathcal{E}^{a}~_{bk}-X^{k}\nabla[\gamma]_{k}g^{ij}\nabla[\gamma]_{i}(A^{a}~_{bj}\\\nonumber-\hat{A}^{a}~_{bj})-X^{k}\nabla[\gamma]_{k}g^{ij}[\hat{A}_{i},A_{j}-\hat{A}_{j}]^{a}~_{b} +g^{ij}\nabla[\gamma]_{i}X^{k}\nabla[\gamma]_{k}A^{a}~_{bj}\\\nonumber+g^{ij}\nabla[\gamma]_{i}A^{a}~_{bk}\nabla[\gamma]_{j}X^{k}+g^{ij}A^{a}~_{bk}\nabla[\gamma]_{i}\nabla[\gamma]_{j}X^{k}-g^{ij}R[\gamma]^{l}_{jik}A^{a}~_{bl}X^{k} \\\nonumber+g^{ij}[\hat{A}_{i},A_{k}]^{a}~_{b}\nabla[\gamma]_{j}X^{k}+(2Nk^{ij}-\nabla^{i}X^{j}-\nabla^{j}X^{i})\nabla[\gamma]^{\hat{A}}_{i}(A^{a}~_{bj}-\hat{A}^{a}~_{bj})=0.
\end{eqnarray}
The modified system coincides with the actual Einstein-Yang-Mills system in the CMCSHGC gauge. If $\mathcal{GA}$ denotes the group of gauge transformations, then the configuration space of the gauge fixed coupled dynamics (`true' dynamics) is identified with the space $\mathcal{M}_{\Sigma}/diff(\Sigma)\times \mathcal{A}/\mathcal{GA}$, where $\mathcal{A}$ is the space of connections on $\Sigma\times \mathbb{R}$. 
This complete coupled elliptic-hyperbolic system is designated as \textbf{CMCSHGC} Einstein-Yang-Mills system. 
\section{Important inequalities}
Here we recall a few important inequalities which are crucial in obtaining the required energy estimates. All the Sobolev spaces are defined on the spatial manifold $\Sigma$ and the corresponding norms are defined in the notations section. Associated constants depend on the background geometry (i.e., on the metric $\gamma$). We will consider $n\geq 2$.\\ 
\textbf{1.} \cite{globalcauchy} \textit{Let $f\in H^{t_{1}},~g\in H^{t_{2}}$ and $k\leq \min(t_{1},t_{2},t_{1}+t_{2}-\frac{n}{2}),~t_{i}\geq 0$ and some $t_{i}>0$, then the following estimate holds
\begin{eqnarray}
\label{eq:product1}
||fg||_{H^{k}}\lesssim ||f||_{H^{t_{1}}}||g||_{H^{t_{2}}}.
\end{eqnarray}
}\\
\textbf{2.}\cite{Taylor} \textit{For $\mathcal{P}\in \mathcal{OP}^{s},~s\in \mathbb{R}$, the following holds
\begin{eqnarray}
||\mathcal{P}u||_{H^{r}}\lesssim ||u||_{H^{s+r}},~~r\in \mathbb{R}.
\end{eqnarray}
Here $\mathcal{OP}^{s}$ denotes the space of  pseudo-differential operators with symbol in Hormander's class $S^{s}_{1,0}$. For example $\nabla[\gamma]\in \mathcal{OP}^{1}$.
}\\
\textbf{3.}\cite{Taylor} \textit{Assume $\mathcal{P}\in \mathcal{OP}^{s},~s>0$, $\sigma\geq 0$ then
\begin{eqnarray}
||[P,u]v||_{H^{\sigma}} \lesssim \left(||\nabla[\gamma] u||_{L^{\infty}}||v||_{H^{s-1+\sigma}}+||u||_{H^{s+\sigma}}||v||_{L^{\infty}}\right)
\end{eqnarray}
and more specifically $[\nabla[\gamma],\nabla[\gamma]^{s-1}]\in \mathcal{OP}^{s-1}$.
}\\
\textbf{4.}(Kato and Ponce \cite{kato1988commutator}) \textit{If $s>0$, then $H^{s}\cap L^{\infty}$ is an algebra i.e., 
\begin{eqnarray}
\label{eq:product2}
||uv||_{H^{s}}\lesssim ||u||_{L^{\infty}}||v||_{H^{s}}+||u||_{H^{s}}||v||_{L^{\infty}}
\end{eqnarray}
and for $s>\frac{n}{2}+1$ due to Sobolev embedding
\begin{eqnarray}
||uv||_{H^{s}}\lesssim ||u||_{H^{s}}||v||_{H^{s}}.
\end{eqnarray}
}
In addition, we will make use of Sobolev embedding, Holder's and Minkowski's inequality whenever necessary.

\section{Energy Estimates}
In the study of hyperbolic PDEs, the use of energy functionals is indispensable in proving local and global well-posedness results. We define the following ad-hoc energies for the Yang-Mills and Einstein sectors. These energies will \textit{not} be gauge invariant and therefore are only defined in our choice of gauge. One could in principle define gauge-invariant energies, but, for the current purpose, this makes no difference. Let us first define a few relevant entities. Let $\Lambda[g]$ be the ellipticity constant of $g$ defined as the least $\Lambda$ such that the following holds 
\begin{eqnarray}
\Lambda^{-1}\gamma(\xi,\xi)\leq g(\xi,\xi)\leq \Lambda\gamma(\xi,\xi),~\forall \xi\in sections\{T\Sigma\}.
\end{eqnarray}
From the expression of the Lorentzian metric $\hat{g}$ in terms of $(g,N,X)$ (\ref{eq:metric}), we define the following entity
\begin{eqnarray}
\Lambda[\hat{g}]:=\Lambda[g]+||N||_{L^{\infty}}+||N^{-1}||_{L^{\infty}}+||X||_{L^{\infty}}
\end{eqnarray}
and for a curve $t\mapsto \hat{g}(t)$, 
\begin{eqnarray}
\Lambda_{t^{*}}:=\sup_{t\in[0,t^{*}]}\Lambda[\hat{g}(t)].
\end{eqnarray}
Using these definitions, calculation yields the following 
\begin{eqnarray}
C(\Lambda[\hat{g}])^{-1}(|\nabla[\gamma]g|+|\nabla N|+|\nabla[\gamma]X|)\leq |\nabla[\gamma]\hat{g}|\\\nonumber 
\leq C(\Lambda[\hat{g}])(|\nabla[\gamma]g|+|\nabla N|+|\nabla[\gamma]X|).
\end{eqnarray}
We define the Yang-Mills energy as follows 
\begin{eqnarray}
E_{YM}:=\frac{1}{2}\left(\int_{\Sigma}A^{a}~_{bi} A^{b}~_{aj}\gamma^{ij}\mu_{g}+\sum_{I=1}^{s}\int_{\Sigma}\nabla[\gamma]^{I}A^{a}~_{bi}\cdot \nabla[\gamma]^{I}A^{b}~_{aj}\gamma^{ij}\mu_{g}\nonumber\right.\\
\left.+\sum_{I=1}^{s}\int_{\Sigma}\nabla[\gamma]^{I-1}\mathcal{E}^{a}~_{bi}\cdot\nabla[\gamma]^{I-1}\mathcal{E}^{b}~_{aj}\gamma^{ij}\mu_{g}\right).
\end{eqnarray}
The energy for the gravitational sector is defined as 
\begin{eqnarray}
E_{Ein}:=\frac{1}{2}\left(\int_{\Sigma}g_{ij}g_{kl}\gamma^{ik}\gamma^{jl}\mu_{g}+\sum_{I=1}^{s}\int_{\Sigma}\nabla[\gamma]^{I}g_{ij}\cdot\nabla[\gamma]^{I}g_{kl}\gamma^{ik}\gamma^{jl}\mu_{g}\nonumber\right.\\
\left. +4\sum_{I=1}^{s}\int_{\Sigma}\nabla[\gamma]^{I-1}k_{ij}\cdot\nabla[\gamma]^{I-1}k_{kl}\gamma^{ik}\gamma^{jl}\mu_{g}\right),
\end{eqnarray}
$s>\frac{n}{2}+1$. In view of the fact that $\Lambda^{-1}\gamma(\xi,\xi)\leq g(\xi,\xi)\leq \Lambda\gamma(\xi,\xi)$, the total energy is equivalent to the desired norm of the data i.e., 
\begin{eqnarray}
E_{s}:=E_{YM}+E_{Ein}\approx ||A||^{2}_{H^{s}}+||\mathcal{E}||^{2}_{H^{s-1}}+||g||^{2}_{H^{s}}+||k||^{2}_{H^{s-1}}.
\end{eqnarray}
We will show that there is unique solution $(g,k,A,\mathcal{E})\in \mathcal{C}([0,t^{*}];H^{s}\times H^{s-1}\times H^{s}\times H^{s-1})$ for $s>\frac{n}{2}+1$ and $t^{*}>0$. The multi index $I$ is defined as follows 
\begin{eqnarray}
\nabla[\gamma]^{I}:=\nabla[\gamma]_{i_{1}}\nabla[\gamma]_{i_{2}}\nabla[\gamma]_{i_{3}}....\nabla[\gamma]_{i_{I}}
\end{eqnarray}
and the dot product
\begin{eqnarray}
\nabla[\gamma]^{I}u\cdot \nabla[\gamma]^{I}v\\
:=g^{i_{1}j_{1}}....g^{i_{I}j_{I}}\nabla[\gamma]_{i_{1}}\nabla[\gamma]_{i_{2}}\nonumber\nabla[\gamma]_{i_{3}}....\nabla[\gamma]_{i_{I}}u\cdot \nabla[\gamma]_{j_{1}}\nabla[\gamma]_{j_{2}}\nabla[\gamma]_{j_{3}}....\nabla[\gamma]_{j_{I}}v,
\end{eqnarray}
where the $\cdot$ indicates the contraction of the remaining indices of $u$ and $v$ with respect to the background metric $\gamma$ and the metric on $\mathfrak{g}$ (if $u$ and $v$ contain internal indices). We start with the first (lowest) order energy. Multiple applications of the operator $\langle \nabla[\gamma]\rangle:=(1+\Delta_{\gamma})^{1/2}$ would produce the desired higher order energy since $W^{s,p}(\Sigma)=\langle \nabla[\gamma]\rangle^{s}L^{p}(\Sigma),~1<p<\infty,~s\in \mathbb{R}$, ($\langle \nabla[\gamma]\rangle^{s}$ acts on the spectral side) with norm defined as 
\begin{eqnarray}
||v||_{W^{s,p}(\Sigma)}=||\langle \nabla[\gamma]\rangle^{s}v||_{L^{p}(\Sigma)}.
\end{eqnarray}
Here $\Delta_{\gamma}$ is defined as $\Delta_{\gamma}:=-\gamma^{ij}\nabla[\gamma]_{i}\nabla[\gamma]_{j}$. If $s$ is a non-negative integer, then $W^{s,p},~1\leq p\leq \infty$ is the closure of $C^{\infty}(\Sigma)$ with respect to the equivalent norm $\sum_{|I|\leq s}||\nabla[\gamma]^{I}v||_{L^{p}(\Sigma)}$. 
The first order energy is defined as follows
\begin{eqnarray}
E_{1}:=E^{1}_{YM}+E^{1}_{Ein}\\
=\frac{1}{2}\left(\int_{\Sigma}A^{a}~_{bi} A^{b}~_{aj}\gamma^{ij}\mu_{g}+\int_{\Sigma}g^{kl}\nabla[\gamma]_{k}A^{a}~_{bi} \nabla[\gamma]_{l}A^{b}~_{aj}\gamma^{ij}\mu_{g}\nonumber\right.\\
\left.+\int_{\Sigma}\mathcal{E}^{a}~_{bi}\mathcal{E}^{b}~_{aj}\gamma^{ij}\mu_{g}\right)\\\nonumber 
+\frac{1}{2}\left(\int_{\Sigma}g_{ij}g_{kl}\gamma^{ik}\gamma^{jl}\mu_{g}+\int_{\Sigma}g^{mn}\nabla[\gamma]_{m}g_{ij}\nabla[\gamma]_{n}g_{kl}\gamma^{ik}\gamma^{jl}\mu_{g}\nonumber\right.\\
\left. +4\int_{\Sigma}k_{ij}k_{kl}\gamma^{ik}\gamma^{jl}\mu_{g}\right)
\end{eqnarray}
The higher order energy $E_{s}$ is equivalent to the first order energy of $\langle\nabla[\gamma]\rangle^{s-1}$ applied on $(g,k,A,\mathcal{E})$ i.e., 
\begin{eqnarray}
E_{s}\approx E_{1}(\langle \nabla[\gamma]\rangle^{s-1}A,\langle \nabla[\gamma]\rangle^{s-2}\mathcal{E}, \langle \nabla[\gamma]\rangle^{s-1}g, \langle \nabla[\gamma]\rangle^{s-2}k),~~s\geq 1.
\end{eqnarray}
One important fact we will note here is that $\langle \nabla[\gamma]\rangle$ is time-independent since the background metric $\gamma$ is time-independent. Let us denote $\langle\nabla[\gamma]\rangle$ by $\langle D\rangle$. Before moving to the energy estimate, let us split the evolution equations into the linear and non-linear parts (in terms of the evolution variables $(g,k,A,\mathcal{E})$) as follows 
\begin{eqnarray}
\mathcal{L}[g,k,N,X]\mathcal{V}\\
=\left[\begin{array}{c}
\partial_{t}g_{ij}+2Nk_{ij}-X^{k}\nabla[\gamma]_{k}g_{ij}\\
\partial_{t}k_{ij}+\frac{N}{2}\Delta_{\gamma}g_{ij}-N\tr_{g}k k_{ij}-X^{k}\nabla[\gamma]_{k}k_{ij}\\
\partial_{t}A^{a}~_{bi}-N\mathcal{E}^{a}~_{bi}-X^{k}\nabla[\gamma]_{k}A^{a}~_{bi}\\
\partial_{t}\mathcal{E}^{a}~_{bi}-Ng^{kl}\nabla[\gamma]_{k}\nabla[\gamma]_{l} A^{a}~_{bi}-N\tr_{g}k\mathcal{E}^{a}~_{bi}-X^{k}\nabla[\gamma]_{k}\mathcal{E}^{a}~_{bi}\nonumber
\end{array}
\right],
\end{eqnarray}
\begin{eqnarray}
\mathcal{V}:=\left[\begin{array}{c}
g\\
k\\
A\\
\mathcal{E}
\end{array}
\right],
\end{eqnarray}
\begin{eqnarray}
\mathcal{F}[g,k,N,X,A,\mathcal{E}]\\
=\left[\begin{array}{c}
g_{ik}\nabla[\gamma]_{j}X^{k}+g_{kj}\nabla[\gamma]_{i}X^{k}\\
\left\{N\mathcal{N}_{ij}-2Nk_{ik}k^{k}_{j}+N\left(\mathcal{E}^{a}~_{bi} \mathcal{E}^{b}~_{aj}-g^{kl}F^{a}~_{bik}\cdot F^{b}~_{ajl}-\frac{g^{kl}\mathcal{E}^{a}~_{bk} \mathcal{E}^{b}_{al}}{n-1}g_{ij}\right.\right.\\\nonumber\left.\left.+\frac{g^{mk}g^{nl}F^{a}~_{bmn}F^{b}~_{akl}}{2(n-1)}g_{ij}\right)-\nabla[\gamma]_{i}\nabla_{j}N\right.\\\left.-\frac{1}{2}g^{kl}(\nabla[\gamma]_{i}g_{lj}\nonumber+\nabla[\gamma]_{j}g_{il}-\nabla[\gamma]_{l}g_{ij})\partial_{k}N+k_{ik}\nabla[\gamma]_{j}X^{k}+k_{kj}\nabla[\gamma]_{i}X^{k}\right\}\\ 
\partial_{i}\varphi^{a}_{b}+[A_{i},\varphi]^{a}~_{b}+A^{a}~_{bj}\nabla[\gamma]_{i}X^{j}\\
N\mathcal{SL}^{a}~_{bi}-F^{a}~_{ai}~^{j}\nabla_{j}N+N[F~_{i}~^{j},A_{j}]^{a}~_{b}+[\mathcal{E}_{i},\varphi]^{a}~_{b}+\mathcal{E}^{a}~_{bk}\nabla[\gamma]_{i}X^{k}-2Nk_{i}^{j}\mathcal{E}^{a}~_{bj}\\\nonumber 
\end{array}
\right]\\
=\left[\begin{array}{c}
\mathcal{F}_{1}\\
\mathcal{F}_{2}\\
(\mathcal{F}_{3})^{a}~_{b}\\
(\mathcal{F}_{4})^{a}~_{b}
\end{array}
\right]
\end{eqnarray}
and therefore the system of evolution equations read 
\begin{eqnarray}
\mathcal{L}[\mathcal{V}]\mathcal{V}=F[\mathcal{V}].
\end{eqnarray}
Let us now derive the necessary energy estimates. First define the following space
\begin{eqnarray}
\mathcal{H}^{s}:=H^{s}(\Sigma)\times H^{s-1}(\Sigma)\times H^{s}(\Sigma)\times H^{s-1}(\Sigma),~~s>\frac{n}{2}+1.
\end{eqnarray}

\textbf{Lemma 1:} \textit{Assume $\mathcal{V}\in L^{\infty}([0,t];\mathcal{H}^{1}\cap \mathcal{W}^{1,\infty})$ solves the EYM evolution equations. Then we have 
\begin{eqnarray}
\sqrt{E_{1}(t)}\leq Ce^{C\int_{0}^{t}(||D\hat{g}(t)||_{L^{\infty}}+||\mathcal{E}||_{L^{\infty}}+||DA||_{L^{\infty}})\nonumber dt^{'}}(\sqrt{E_{1}(0)}+||\mathcal{F}||_{L^{1}([0,t];\mathcal{H}^{1})})
\end{eqnarray}
for a constant $C=C(\Lambda_{t})$.}\\
\textbf{Proof:}
We now explicitly evaluate the time evolution of the first order energy
\begin{eqnarray}
\frac{dE_{1}}{dt}=\int_{\Sigma}\left(\gamma^{ij}(\partial_{t}A^{a}~_{bi})A^{b}~_{aj}+g^{ij}\nabla[\gamma]_{i}\partial_{t}A^{a}~_{bm}\nabla[\gamma]_{j}A^{b}~_{an}\gamma^{mn}\nonumber\right.\\\nonumber 
\left.
-\frac{1}{2}g^{ik}g^{jl}\partial_{t}g_{kl}\nabla[\gamma]_{i}A^{a}A_{bm}\nabla[\gamma]_{j}A^{b}~_{an}\gamma^{mn}+\gamma^{ij}\partial_{t}\mathcal{E}^{a}~_{bi}\mathcal{E}^{b}~_{aj}\right.\\\nonumber 
\left.+\gamma^{ik}\gamma^{jl}\partial_{t}g_{ij}g_{kl}+g^{ij}\nabla[\gamma]_{i}\partial_{t}g_{kl}\nabla[\gamma]_{j}g_{mn}\gamma^{km}\gamma^{ln}+4\gamma^{ik}\gamma^{jl}\partial_{t}k_{ij}k_{kl}\right.\\\nonumber 
\left.-\frac{1}{2}g^{ir}g^{js}\partial_{t}g_{rs}\nabla[\gamma]_{i}g_{kl}\nabla[\gamma]_{j}g_{mn}\gamma^{km}\gamma^{ln}\right)\mu_{g}\\\nonumber 
+\frac{1}{4}\int_{\Sigma}\left(A^{a}~_{bi} A^{b}~_{aj}\gamma^{ij}+g^{kl}\nabla[\gamma]_{k}A^{a}~_{bi} \nabla[\gamma]_{l}A^{b}~_{aj}\gamma^{ij}\nonumber\right.\\
\left.+\mathcal{E}^{a}~_{bi}\mathcal{E}^{b}~_{aj}\gamma^{ij}+g_{ij}g_{kl}\gamma^{ik}\gamma^{jl}\mu_{g}+g^{mn}\nabla[\gamma]_{m}g_{ij}\nabla[\gamma]_{n}g_{kl}\gamma^{ik}\gamma^{jl}\nonumber\right.\\
\left. +4k_{ij}k_{kl}\gamma^{ik}\gamma^{jl}\mu_{g}\right)g^{pq}\partial_{t}g_{pq}\mu_{g}.
\end{eqnarray}
Now we use the evolution equations for $\mathcal{V}:=(g,k,A,\mathcal{E})$. A direct calculation yields 
\begin{eqnarray}
\frac{dE_{1}}{dt}=\int_{\Sigma}\left(\gamma^{ij}(N\mathcal{E}^{a}~_{bi}+X^{k}\nabla[\gamma]_{k}A^{a}~_{bi}\nonumber+(\mathcal{F}~_{3})^{a}~_{bi})A^{b}~_{aj}\right.\\\nonumber
\left.+g^{ij}\nabla[\gamma]_{i}(N\mathcal{E}^{a}~_{bm}+X^{k}\nabla[\gamma]_{k}A^{a}~_{bm}+(\mathcal{F}_{3})^{a}~_{bm})\nabla[\gamma]_{j}A^{a}~_{bn}\gamma^{mn}\right.\\\nonumber
\left.+\gamma^{ij}(N\Delta^{\gamma}_{g}A^{a}~_{bi}+N\tr_{g}k \mathcal{E}^{a}~_{bi}+X^{k}\nabla[\gamma]_{k}\mathcal{E}^{a}~_{bi}+(\mathcal{F}_{4})_{i})\mathcal{E}^{a}~_{bj}\right.\\\nonumber
\left.+\gamma^{ik}\gamma^{jl}(-2Nk_{ij}+X^{k}\nabla[\gamma]_{k}g_{ij}+(\mathcal{F}_{1})_{ij})g_{kl}\right.\\\nonumber
\left. +g^{ij}\nabla[\gamma]_{i}(-2Nk_{kl}+X^{m}\nabla[\gamma]_{m}g_{kl}+(\mathcal{F}_{1})_{kl})\nabla[\gamma]_{j}g_{mn}\gamma^{km}\gamma^{ln}\right.\\\nonumber
\left.+4\gamma^{ik}\gamma^{jl}(-\frac{N}{2}\Delta^{\gamma}_{g}g_{ij}+N\tr_{g}k k_{ij}+X^{k}\nabla[\gamma]_{k}k_{ij}+(\mathcal{F}_{2})_{ij})k_{kl}\right.\\\nonumber 
\left.-\frac{1}{2}g^{ik}g^{jl}(-2Nk_{ij}+X^{m}\nabla[\gamma]_{m}g_{kl}+(\mathcal{F}_{1})_{kl})\nabla[\gamma]_{i}A^{a}~_{bm}\nabla[\gamma]_{j}A^{b}~_{an}\gamma^{mn}\right.\\\nonumber 
\left.-\frac{1}{2}g^{ir}g^{js}(-2Nk_{rs}+X^{m}\nabla[\gamma]_{m}g_{rs}+(\mathcal{F}_{1})_{rs})\nabla[\gamma]_{i}g_{kl}\nabla[\gamma]_{j}g_{mn}\gamma^{km}\gamma^{ln}\right)\mu_{g}\\\nonumber 
+\frac{1}{4}\int_{\Sigma}\left(A^{a}~_{bi} A^{b}~_{aj}\gamma^{ij}+g^{kl}\nabla[\gamma]_{k}A^{a}~_{bi} \nabla[\gamma]_{l}A^{b}~_{aj}\gamma^{ij}\nonumber\right.\\
\left.+\mathcal{E}^{a}~_{bi}\mathcal{E}^{b}~_{aj}\gamma^{ij}+g_{ij}g_{kl}\gamma^{ik}\gamma^{jl}\mu_{g}+g^{mn}\nabla[\gamma]_{m}g_{ij}\nabla[\gamma]_{n}g_{kl}\gamma^{ik}\gamma^{jl}\nonumber\right.\\
\left. +4k_{ij}k_{kl}\gamma^{ik}\gamma^{jl}\mu_{g}\right)(-2N\tr_{g}k+\nabla_{k}X^{k})\mu_{g}
\end{eqnarray}
We notice that the principal terms of the type $\Delta^{\gamma}_{g}A^{a}~_{bi}\mathcal{E}^{b}~_{aj}\gamma^{ij}$ must cancel with their negative counter parts point-wise in order for the energy argument to close. We compute
\begin{eqnarray}
\gamma^{ij}g^{kl}(\nabla[\gamma]_{k}\nabla[\gamma]_{l}A^{a}~_{bi})(N\mathcal{E}^{b}~_{aj})\\
=\underbrace{\nabla[\gamma]_{k}(\gamma^{ij}g^{kl}N\mathcal{E}^{a}~_{bj}\nabla[\gamma]_{l}A^{b}~_{ai})}_{I: vanishes~by~integration}\nonumber-\underbrace{\gamma^{ij}g^{kl}\nabla[\gamma]_{k}(N\mathcal{E}^{a}~_{bj})\nabla[\gamma]_{l}A^{b}~_{ai}}_{II}\\\nonumber 
-\gamma^{ij}N\mathcal{E}^{a}~_{bj}\nabla[\gamma]_{k}g^{kl}\nabla[\gamma]_{l}A^{b}~_{ai}
\end{eqnarray}
and notice that the first term vanishes while integrated since it is a total derivative while the second term $II$ cancels out point-wise with its negative. The remaining term is harmless in terms of regularity. An exactly similar calculation holds for the principal term arising from the gravitational sector. In addition, we also have terms that contain one extra derivative (such as $g^{ij}\nabla[\gamma]_{i}(X^{k}\nabla[\gamma]_{k}A^{a}~_{bm})$, $\gamma^{ik}\gamma^{jl}X^{m}\nabla[\gamma]_{m}k_{ij}k_{kl}$ etc) which are potential obstructions to closing the energy argument. However, these terms may be controlled by manipulating the derivatives. For example, compute
\begin{eqnarray}
\int_{\Sigma}\gamma^{ik}\gamma^{jl}X^{m}\nabla[\gamma]_{m}k_{ij}k_{kl}\mu_{g}=\frac{1}{2}\int_{\Sigma}X^{m}\nabla[\gamma]_{m}(\gamma^{ik}\gamma^{jl}k_{ij}k_{kl})\mu_{g}\\\nonumber 
=\frac{1}{2}\int_{\Sigma}\nabla[\gamma]_{m}\left(X^{m}\gamma^{ik}\gamma^{jl}k_{ij}k_{kl}\mu_{g}\right)-\frac{1}{2}\int_{\Sigma}\nabla[\gamma]_{m}X^{m}\gamma^{ik}\gamma^{jl}k_{ij}k_{kl}\mu_{g}\\\nonumber 
-\frac{1}{2}\int_{\Sigma}X^{m}\gamma^{ik}\gamma^{jl}k_{ij}k_{kl}(\Gamma[g]^{k}_{km}-\Gamma[\gamma]^{k}_{km})\mu_{g}\\
=-\frac{1}{2}\int_{\Sigma}\nabla[\gamma]_{m}X^{m}\gamma^{ik}\gamma^{jl}k_{ij}k_{kl}\mu_{g}-\frac{1}{2}\int_{\Sigma}X^{m}\gamma^{ik}\gamma^{jl}k_{ij}k_{kl}(\Gamma[g]^{k}_{km}\nonumber-\Gamma[\gamma]^{k}_{km})\mu_{g},
\end{eqnarray}
and therefore 
\begin{eqnarray}
|\int_{\Sigma}\gamma^{ik}\gamma^{jl}X^{m}\nabla[\gamma]_{m}k_{ij}k_{kl}\mu_{g}|
\\\nonumber \lesssim ||\nabla[\gamma]X||_{L^{\infty}}\int_{\Sigma}|k|^{2}_{\gamma}\mu_{g}+||k||_{L^{\infty}}||X||_{L^{\infty}}\int_{\Sigma}\left(|k|^{2}_{\gamma}+|\nabla[\gamma]g|^{2}_{\gamma}\right)\mu_{g}.
\end{eqnarray}
Other terms of this type may be handled in a similar way. Collecting all the terms and applying Holder's inequality, we obtain the following  differential inequality for the energy
\begin{eqnarray}
\frac{dE_{1}}{dt}\leq C(\Lambda_{t})\left(E_{1}\nonumber+\nabla[\gamma]g||_{L^{\infty}}E_{1}+|\tr_{g}k|E_{1}+||\nabla[\gamma]g||_{L^{\infty}}E_{1}\right.\\\nonumber
\left.+||\nabla[\gamma]X||_{L^{\infty}}E_{1}+||Riem[\gamma]||_{L^{\infty}}E_{1}\nonumber+||\nabla[\gamma]A||_{L^{\infty}}E_{1}+||\mathcal{E}||_{L^{\infty}}E_{1}\right.\\\nonumber 
\left.+||k||_{L^{\infty}}E_{1}+||\nabla[\gamma]g||_{L^{\infty}}E_{1}+||\mathcal{F}_{3}||_{L^{2}}\sqrt{E_{1}}+||\mathcal{F}_{1}||_{H^{1}}\sqrt{E_{1}}\right.\\\nonumber 
\left.+||\mathcal{F}_{4}||_{L^{2}}\sqrt{E_{1}}+||\mathcal{F}_{1}||_{L^{2}}\sqrt{E_{1}}+||\mathcal{F}_{2}||_{L^{2}}\sqrt{E_{1}}\right)
\end{eqnarray}
integration of which yields
\begin{eqnarray}
\sqrt{E_{1}(t)}-\sqrt{E_{1}(0)}\leq C(\Lambda_{t}) \int_{0}^{t}\left((1+||\nabla[\gamma]g||_{L^{\infty}}\nonumber+|\tr_{g}k|)\sqrt{E}_{1}(t^{'})\right.\\\nonumber 
\left.+(||k||_{L^{\infty}}+||Riem[\gamma]||_{L^{\infty}}+||\nabla[\gamma]g||_{L^{\infty}})\sqrt{E(t^{'})}+(||\nabla[\gamma]X||_{L^{\infty}}\right.\\\nonumber \left.+||\nabla[\gamma]A||_{L^{\infty}}+||\mathcal{E}||_{L^{\infty}})\sqrt{E(t^{'})}\right)dt^{'}+C(\Lambda_{t})||\mathcal{F}||_{L^{1}([0,t];\mathcal{H}^{1})}.
\end{eqnarray}
Noting $\gamma$ is a $C^{\infty}$ background metric, $||Riem[\gamma]||_{L^{\infty}}\lesssim 1$ and applying Gr\"onwall's inequality, we obtain 
\begin{eqnarray}
\sqrt{E_{1}(t)}\leq \left(\sqrt{E_{1}(0)}\nonumber+C(\Lambda_{t})||\mathcal{F}||_{L^{1}([0,t];\mathcal{H}^{1})}\right)e^{C(\Lambda_{t})\int_{0}^{t}\mathcal{B}(t^{'})dt^{'}},
\end{eqnarray}
where $\mathcal{B}(t)$ is defined as follows 
\begin{eqnarray}
\mathcal{B}(t)&:=&\left(||\nabla[\gamma]X||_{L^{\infty}}\nonumber+||\nabla[\gamma]g||_{L^{\infty}}+||k||_{L^{\infty}}+||\mathcal{E}||_{L^{\infty}}+||\nabla[\gamma]A||_{L^{\infty}}\right)\\\nonumber
&\leq& (||D\hat{g}(t)||_{L^{\infty}}+||\mathcal{E}||_{L^{\infty}}+||DA||_{L^{\infty}}).~~~~~~~~~~~~~~~~~~~~~~~~~~~~~~~~~\square
\end{eqnarray}
Now in the view of Sobolev embedding for $s>\frac{n}{2}+1$, $\mathcal{B}$ is dominated by the available Sobolev norm of $(g,k,X,A,\mathcal{E})$. We need to obtain higher order energy estimates.\\ 
\textbf{Lemma 2:} \textit{Assume $\mathcal{V}\in L^{\infty}([0,t];\mathcal{H}^{s}\cap \mathcal{W}^{1,\infty}),~s>\frac{n}{2}+1$ solves the EYM evolution equations. Then we have 
\begin{eqnarray}
\sqrt{E_{s}(t)}&\leq& Ce^{C(\Lambda_{t})\int_{0}^{t}(||\hat{g}||_{H^{s}}+||k||_{H^{s-1}}+||A||_{H^{s}}+||\mathcal{E}||_{H^{s-1}})dt^{'}}\\&&(\sqrt{E_{s}(0)}+||\mathcal{F}||_{L^{1}([0,t];\mathcal{H}^{s})}),
\end{eqnarray}
where $||\hat{g}||_{H^{s}}\approx ||g||_{H^{s}}+||N||_{H^{s+1}}+||X||_{H^{s+1}}$
}.

\textbf{Proof:}
Now we need to prove a higher order estimate. We want to estimate the energy corresponding to $\langle D\rangle ^{s-1}\mathcal{V}$. Use the commutator 
\begin{eqnarray}
\mathcal{L}\langle D\rangle ^{s-1}\mathcal{V}=[\mathcal{L},\langle D\rangle^{s-1}]\mathcal{V}+\langle \mathcal{D}\rangle ^{s-1}\mathcal{F}
\end{eqnarray}
Using the energy estimate  of lemma $1$, one obtains 
\begin{eqnarray}
\sqrt{E_{s}(t)}&\leq& Ce^{\int_{0}^{t}C(||D\hat{g}(t)||_{L^{\infty}}+||\mathcal{E}||_{L^{\infty}}+||DA||_{L^{\infty}})dt^{'}}\\\nonumber
&&(\sqrt{E_{s}(0)}+||\langle D\rangle^{s-1}\mathcal{F}||_{L^{1}([0,t];\mathcal{H}^{1})}
+||[L,\langle D\rangle^{s-1}]\mathcal{V}||_{L^{1}([0,t];\mathcal{H}^{1})})
\end{eqnarray}
Now we have to show that the energy argument closes, that is, $||[L,\langle D\rangle^{s-1}]\mathcal{V}||_{\mathcal{H}^{1}}$ can be estimated using the maximum available Sobolev norms. We will use the inequalities listed in the previous section (section 4). 
\begin{eqnarray}
[\partial_{t},\langle D\rangle^{s-1}]=0
\end{eqnarray}
\begin{eqnarray}
||[N,\langle D\rangle ^{s-1}]k||_{H^{1}}\leq C\left(||\nabla[\gamma]N||_{L^{\infty}}||k||_{H^{s-1}}+||N||_{H^{s}}||k||_{L^{\infty}}\right)
\end{eqnarray}
\begin{eqnarray}
[\langle D\rangle^{s-1},X^{k}\nabla[\gamma]_{k}]g=[\langle D\rangle^{s-1},X^{k}]\nabla[\gamma]_{k}g+X^{k}[\langle D\rangle^{s-1},\nabla[\gamma]_{k}]g
\end{eqnarray}
Note $[\langle D\rangle^{s-1},\nabla[\gamma]_{k}]\in \mathcal{OP}^{s-1}$ and following inequality (\ref{eq:product1})
\begin{eqnarray}
||[\langle D\rangle^{s-1},X^{k}\nabla[\gamma]_{k}]g||_{H^{1}}\\\nonumber 
\leq ||[\langle D\rangle^{s-1},X^{k}]\nabla[\gamma]_{k}g||_{H^{1}}\nonumber+||X^{k}[\langle D\rangle^{s-1},\nabla[\gamma]_{k}]g||_{H^{1}}\\\nonumber 
\lesssim ||\nabla[\gamma] X||_{L^{\infty}}||g||_{H^{s}}+||X||_{H^{s}}||\nabla[\gamma]g||_{L^{\infty}}.
\end{eqnarray}
We need to estimate the term of the following type
\begin{eqnarray}
||[N\Delta^{\gamma}_{g},\langle D\rangle ^{s-1}]g||_{L^{2}}
\end{eqnarray}
where 
\begin{eqnarray}
\Delta^{\gamma}_{g}:=g^{ij}\nabla[\gamma]_{i}\nabla[\gamma]_{j}
\end{eqnarray}
is of the type $g^{-1}D^{2}$. We perform a series of manipulations of the commutators 
\begin{eqnarray}
[N\Delta_{\gamma},\langle D\rangle ^{s-1}]g=
[Ng^{-1}D^{2},\langle D\rangle ^{s-1}]g\\\nonumber 
=Ng^{-1}[D^{2},\langle D\rangle ^{s-1}]g+\underbrace{[Ng^{-1},\langle D\rangle ^{s-1}]D^{2}g}_{II}
\end{eqnarray}
Now we notice that the second term $II$ is potentially dangerous and therefore requires further attention. We further simplify it as follows 
\begin{eqnarray}
[Ng^{-1},\langle D\rangle ^{s-1}]D^{2}g=Ng^{-1}\langle D\rangle ^{s-1}D^{2}g-\langle D\rangle^{s-1}(Ng^{-1}D^{2}g)\\
=Ng^{-1}\langle D\rangle ^{s-1}D Dg-\langle D\rangle ^{s-1}\left(D(Ng^{-1}Dg)-D(Ng^{-1})Dg\right)\\
=[Ng^{-1},\langle D\rangle^{s-1}D]Dg+\langle D\rangle ^{s-1}\left(D(Ng^{-1})Dg\right)
\end{eqnarray}
and therefore see that all the terms are estimated by the maximum available Sobolev norms. Using the commutator estimate, we obtain 
\begin{eqnarray}
||[N\Delta_{\gamma},\langle D\rangle ^{s-1}]g||_{L^{2}}\lesssim ||N||_{H^{s}}||\nabla[\gamma]g||_{L^{\infty}}+||\nabla[\gamma]N||_{L^{\infty}}||g||_{H^{s}}.
\end{eqnarray}
The remaining terms are estimated similarly by means of the commutator estimate 
\begin{eqnarray}
||[\langle D\rangle^{s-1},X^{k}\nabla[\gamma]_{k}]k||_{L^{2}}\lesssim ||X||_{L^{\infty}}||k||_{H^{s-1}}+||X||_{H^{s}}||k||_{L^{\infty}},\\
||[N,\langle D\rangle ^{s-1}]\mathcal{E}||_{H^{1}}\lesssim \left(||\nabla[\gamma]N||_{L^{\infty}}||\mathcal{E}||_{H^{s-1}}+||N||_{H^{s}}||\mathcal{E}||_{L^{\infty}}\right),\\
||[\langle D\rangle^{s-1},X^{k}\nabla[\gamma]_{k}]A||_{H^{1}}\lesssim ||\nabla[\gamma] X||_{L^{\infty}}||A||_{H^{s}}+||X||_{H^{s}}||\nabla[\gamma]A||_{L^{\infty}},\\
||[N\Delta_{\gamma},\langle D\rangle ^{s-1}]A||_{L^{2}}\lesssim ||N||_{H^{s}}||\nabla[\gamma]A||_{L^{\infty}}+||\nabla[\gamma]N||_{L^{\infty}}||A||_{H^{s}},\\
||[\langle D\rangle^{s-1},X^{k}\nabla[\gamma]_{k}]\mathcal{E}||_{L^{2}}\lesssim ||X||_{L^{\infty}}||\mathcal{E}||_{H^{s-1}}+||X||_{H^{s}}||\mathcal{E}||_{L^{\infty}}
\end{eqnarray}
Consider the equation 
\begin{eqnarray}
L\langle D\rangle ^{s-1}\mathcal{V}=[L,\langle D\rangle^{s-1}]\mathcal{V}+\langle \mathcal{D}\rangle ^{s-1}\mathcal{F}
\end{eqnarray}
and apply the energy inequality established in the previous lemma to obtain 
\begin{eqnarray}
\sqrt{E_{s}(t)}-\sqrt{E_{s}(0)}\leq C(\Lambda_{t})\int_{0}^{t}\left((1+||\nabla[\gamma]g||_{L^{\infty}}\nonumber+|\tr_{g}k|)\sqrt{E}_{s}(t^{'})\right.\\\nonumber 
\left.+(||k||_{L^{\infty}}+||Riem[\gamma]||_{L^{\infty}}+||\mathcal{E}||_{L^{\infty}}+||\nabla[\gamma]A||_{L^{\infty}}+||\nabla[\gamma]g||_{L^{\infty}})\sqrt{E_{s}(t^{'})}\right.\\\nonumber \left.+||\nabla[\gamma]X||_{L^{\infty}}\sqrt{E_{s}(t^{'})}\right)dt^{'}+C(\Lambda_{t})||\mathcal{F}||_{L^{1}([0,t];\mathcal{H}^{s})}+C(\Lambda_{t})\int_{0}^{t}||[L,\langle D\rangle^{s-1}]\mathcal{V}(t^{'})||_{\mathcal{H}^{1}}dt^{'}.
\end{eqnarray}
Here we have used the fact that $||\langle D\rangle^{s-1}\mathcal{F}(t)||_{\mathcal{H}^{1}}\leq C(\Lambda_{t})||\mathcal{F}(t)||_{\mathcal{H}^{s}}$.
Now, using the previous calculations, we have
\begin{eqnarray}
\int_{0}^{t}||[L,\langle D\rangle^{s-1}]\mathcal{V}(t^{'})||_{\mathcal{H}^{1}}dt^{'}\leq C(\Lambda_{t})\int_{0}^{t}\nonumber||D\hat{g}||_{L^{\infty}}||\mathcal{V}(t^{'})||_{\mathcal{H}^{s}}dt^{'}\\\nonumber 
\leq C(\Lambda_{t})\int_{0}^{t}(\nonumber||D\hat{g}||_{L^{\infty}})\sqrt{E_{s}(t^{'})}dt^{'}
\end{eqnarray}
and therefore 
\begin{eqnarray}
\sqrt{E_{s}(t)}-\sqrt{E_{s}(0)}\leq \int_{0}^{t}C(\Lambda_{t})\left((1+||\nabla[\gamma]g||_{L^{\infty}}\nonumber+|\tr_{g}k|)\sqrt{E}_{s}(t^{'})\right.\\\nonumber 
\left.+(||k||_{L^{\infty}}+||\nabla[\gamma]g||_{L^{\infty}})\sqrt{E_{s}(t^{'})}+(||\nabla[\gamma]X||_{L^{\infty}}+||\nabla[\gamma]g||_{L^{\infty}}\right.\\\nonumber \left.+||k||_{L^{\infty}})\sqrt{E_{s}(t^{'})}\right)dt^{'}+C(\Lambda_{t})||\mathcal{F}||_{L^{1}([0,t];\mathcal{H}^{s})}\\\nonumber 
+C(\Lambda_{t})\int_{0}^{t}\nonumber||D\hat{g}||_{L^{\infty}}\sqrt{E_{s}(t^{'})}dt^{'}.
\end{eqnarray}
Using Gronwall's inequality and Sobolev embedding with $s>\frac{n}{2}+1$ yields the desired estimate (notice that the energy argument closes using Sobolev embedding)
\begin{eqnarray}
\sqrt{E_{s}(t)}&\leq& Ce^{C(\Lambda_{t})\int_{0}^{t}(||\hat{g}||_{H^{s}}+||k||_{H^{s-1}}+||A||_{H^{s}}+||\mathcal{E}||_{H^{s-1}})dt^{'}}\\&&(\sqrt{E_{s}(0)}+||\mathcal{F}||_{L^{1}([0,t];\mathcal{H}^{s})}).\nonumber ~~~~~~~~~~~~~~~~~~~~~~~~~~~~~~~~~~~~~~~~~~~\square
\end{eqnarray}

Let us now perform a bit more analysis of the source term $\mathcal{F}$. We have to show that $||\mathcal{F}||_{L^{1}([0,t];\mathcal{H}^{s})}$ can be controlled by the maximum available Sobolev norms. 
We claim that the non-linear functional $\mathcal{F}$ is locally Lipschitz as the following map
\begin{eqnarray}
\mathcal{F}:\mathcal{H}^{s}\to \mathcal{H}^{s}
\end{eqnarray}
provided that $(N,X,\varphi)\in H^{s+1}(\Sigma)\times H^{s+1}(\Sigma)\times H^{s+1}(\Sigma)$. First we show that the energy argument closes i.e., if $\mathcal{V}\in \mathcal{H}^{s}$, then $\mathcal{F}(\mathcal{V})\in \mathcal{H}^{s}$. From this calculation, the local Lipschitz property of the map $\mathcal{F}:\mathcal{H}^{s}\to\mathcal{H}^{s}$ follows.\\ 
\textbf{Lemma 3:} \textit{Let $\mathcal{V}:=(g,k,A,\mathcal{E})\in \mathcal{H}^{s}$. If $N,X,\varphi\in H^{s+1}\times H^{s+1}\times H^{s+1},~s>\frac{n}{2}+1$, then $\mathcal{F}$ as a map 
\begin{eqnarray}
\mathcal{F}:\mathcal{H}^{s}\to\mathcal{H}^{s},\\\nonumber
\mathcal{V}\mapsto \mathcal{F}(\mathcal{V})
\end{eqnarray}
is continuous.
}

\textbf{Proof:} Using the expression of $\mathcal{F}(\mathcal{V})$ in conjunction with Minkowski's inequality and the algebra property of the Sobolev space $H^{k}$ for $k>\frac{n}{2}$, we obtain 
\begin{eqnarray}
||\mathcal{F}_{1}||_{H^{s}}\leq C||g||_{H^{s}}||X||_{H^{s+1}},\\
||\mathcal{F}_{2}||_{H^{s-1}}\leq C\left(||N||_{H^{s-1}}||\mathcal{N}||_{H^{s-1}}\nonumber+||N||_{H^{s-1}}||k||^{2}_{H^{s-1}}+||N||_{H^{s-1}}||\mathcal{E}||^{2}_{H^{s-1}}\right.\\\nonumber
\left.+||N||_{H^{s-1}}||A||^{2}_{H^{s-1}}+||N||_{H^{s+1}}+||N||_{H^{s}}||g||_{H^{s}}+||k||_{H^{s-1}}||X||_{H^{s}}\right).
\end{eqnarray}
Now from the explicit expression of $\mathcal{N}_{ij}$ (\ref{eq:riccinonlinear}), we obtain 
\begin{eqnarray}
||\mathcal{N}||_{H^{s-1}}\lesssim ||g||^{2}_{H^{s}}
\end{eqnarray}
and therefore
\begin{eqnarray}
||\mathcal{F}_{2}||_{H^{s-1}}\leq C\left(||N||_{H^{s-1}}||g||^{2}_{H^{s}}\nonumber+||N||_{H^{s-1}}||k||^{2}_{H^{s-1}}+||N||_{H^{s-1}}||\mathcal{E}||^{2}_{H^{s-1}}\right.\\\nonumber
\left.+||N||_{H^{s-1}}||A||^{2}_{H^{s-1}}+||N||_{H^{s+1}}+||N||_{H^{s}}||g||_{H^{s}}+||k||_{H^{s-1}}||X||_{H^{s}}\right).
\end{eqnarray}
The remaining components of $\mathcal{F}$ satisfy
\begin{eqnarray}
||\mathcal{F}_{3}||_{H^{s}}\leq C\left(||\varphi||_{H^{s+1}}+||A||_{H^{s}}||\varphi||_{H^{s}}+||A||_{H^{s}}||X||_{H^{s+1}}\right),\\
||\mathcal{F}_{4}||_{H^{s-1}}\leq C\left(||\mathcal{E}||_{H^{s-1}}||\varphi||_{H^{s-1}}+||\mathcal{E}||_{H^{s-1}}||X||_{H^{s}}\nonumber+||N||_{H^{s-1}}||\mathcal{SL}||_{H^{s-1}}\right.\\\nonumber 
\left.+||N||_{H^{s-1}}||k||_{H^{s-1}}||\mathcal{E}||_{H^{s-1}}+||N||_{H^{s-1}}||F||_{H^{s-1}}+||N||_{H^{s-1}}||F||_{H^{s-1}}||A||_{H^{s-1}}\right).
\end{eqnarray}
Once again we need to make sure that the nonlinear term $\mathcal{SL}$ is under control. Recalling the expression of $\mathcal{SL}$ (\ref{eq:YMNL}), we obtain
\begin{eqnarray}
||\mathcal{SL}||_{H^{s-1}}\lesssim ||g||_{H^{s}}||\hat{A}||_{H^{s}}\nonumber+||\hat{A}||_{H^{s+1}}+||g||_{H^{s}}||\hat{A}||_{H^{s-1}}||A-\hat{A}||_{H^{s-1}}\\\nonumber
+||\hat{A}||_{H^{s}}||A-\hat{A}||_{H^{s-1}}+||\hat{A}||_{H^{s-1}}||A-\hat{A}||_{H^{s}}+||g||_{H^{s}}||A||_{H^{s}}\\\nonumber +||R[\gamma]||_{H^{s-1}}||A||_{H^{s-1}}+||g||_{H^{s}}||A||^{2}_{H^{s-1}}+||A||_{H^{s}}||A||_{H^{s-1}}+||g||_{H^{s}}||F||_{H^{s-1}}.
\end{eqnarray}
Now note that $\hat{A}\in C^{\infty}(\mathfrak{X}(\Sigma))$, $\gamma\in C^{\infty}(\mathcal{M}_{\Sigma})$, and $||F||_{H^{s-1}}\lesssim ||A||_{H^{s}}+||A||^{2}_{H^{s-1}}$. Since $N,X,\varphi\in H^{s+1}$, this yields $||\mathcal{F}(\mathcal{V})||_{\mathcal{H}^{s}}\lesssim ||\mathcal{V}||_{\mathcal{H}^{s}}$ which establishes continuity of $\mathcal{F}$.\\
\textbf{Lemma 4:} \textit{Let $\mathcal{V}:=(g,k,A,\mathcal{E})\in \mathcal{H}^{s}$. If $N,X,\varphi\in H^{s+1}\times H^{s+1}\times H^{s+1},~s>\frac{n}{2}+1$, then $\mathcal{F}$ as a map 
\begin{eqnarray}
\mathcal{F}:\mathcal{H}^{k}\to\mathcal{H}^{k},\\\nonumber
\mathcal{V}\mapsto \mathcal{F}(\mathcal{V})
\end{eqnarray}
is locally Lipschitz for $1\leq k\leq s$.
}\\
\textbf{Proof:} We recall the following product estimate (4.1) 
\begin{eqnarray}
||fg||_{H^{k}}\lesssim ||f||_{H^{t_{1}}}||g||_{H^{t_{2}}}
\end{eqnarray}
for $f\in H^{t_{1}},~g\in H^{t_{2}}$ and $k\leq \min(t_{1},t_{2},t_{1}+t_{2}-\frac{n}{2}),~t_{i}\geq 0$ and some $t_{i}>0$.This estimate yields that the multiplication maps $H^{s}\times H^{1}\to H^{1}$ and $H^{s}\times L^{2}\to L^{2}$ are continuous for $s>\frac{n}{2}+1$. Now repeating the calculation exactly as the lemma 3 and using this product estimate, we obtain 
\begin{eqnarray}
||\mathcal{F}(\mathcal{V}_{1})-\mathcal{F}(\mathcal{V}_{2})||_{\mathcal{H}^{k}}\\\nonumber 
\leq C(1+||\mathcal{V}_{1}||_{\mathcal{H}^{s}}+||\mathcal{V}_{2}||_{H^{s}}+||\mathcal{V}_{1}||^{2}_{\mathcal{H}^{s}}\nonumber+||\mathcal{V}_{2}||^{2}_{\mathcal{H}^{s}})||\mathcal{V}_{1}-\mathcal{V}_{2}||_{\mathcal{H}^{k}}.
\end{eqnarray}
Therefore the map $\mathcal{F}:\mathcal{H}^{k}\to \mathcal{H}^{k}$ is also Lipschitz on $\mathcal{B}^{k}_{1/C_{L}}(\mathcal{V}^{0})$ with Lipschitz constant $C_{L}$, that is, there exists a constant $C_{L}$ such that the following holds
\begin{eqnarray}
||\mathcal{F}(\mathcal{V}_{1})-\mathcal{F}(\mathcal{V}_{2})||_{\mathcal{H}^{k}}\leq C_{L}||\mathcal{V}_{1}-\mathcal{V}_{2}||_{\mathcal{H}^{k}}
\end{eqnarray}
for $1\leq k\leq s.$~~~~~~~~~~~~~~~~~~~~~~~~~~~~~~~~~~~~~~~~~~~~~~~~~~~~~~~~~~~~~~~~~~~~~~~~~~~~~~~~~~~~~~~~~~~~$\square$

\subsection{Existence of a Solution}
In this section, we will establish the existence of a unique solution of the gauge fixed Einstein-Yang-Mills equations in the function space $\mathcal{C}\left([0,t^{*}];\mathcal{H}^{s}\right),~s>\frac{n}{2}+1$ for a suitable $t^{*}>0$ utilizing the technique developed by Anderson and Moncrief \cite{andersson2003elliptic}. Their technique is capable of handling the gauge fixed Einstein equations where the gauge group is the diffeomorphism group naturally arising from gravity. Leaving out a few subtleties discussed in the overview section, the major difference between the work of \cite{andersson2003elliptic} and the present article is the inclusion of Yang-Mills source terms. 

Recall that we have data on the initial hypersurface at $t=0$ and given by $\mathcal{V}^{0}:=\mathcal{V}(0)=[g(0),k(0),A(0),\mathcal{E}(0)]^{T}$. Let us now construct a sequence $\mathcal{V}^{0}_{k}\in C^{\infty}\cap B^{s}_{R}(\mathcal{V}^{0}),~k\geq 1$ by applying an approximation to the identity on the initial data $\mathcal{V}^{0}$ i.e., 
\begin{eqnarray}
\lim_{k\to\infty}||\mathcal{V}^{0}-\mathcal{V}^{0}_{k}||_{\mathcal{H}^{s}}=0.
\end{eqnarray}
Here $B^{s}_{R}(\mathcal{V}^{0})$ is a ball of radius $R$ centered at $\mathcal{V}^{0}$ in $\mathcal{H}^{s}$. Since $C^{\infty}$ is dense in Sobolev spaces, this mollification procedure is standard. Now we construct a sequence of approximate solutions $\{\mathcal{V}_{k}\}_{k=1}^{\infty}\in L^{\infty}([0,t^{*}];B^{s}_{R}(\mathcal{V}^{0}))$ with initial data given by $\mathcal{V}^{0}_{k}$ and prove that this sequence converges in $\mathcal{C}([0,t^{*}];B^{s}_{R}(\mathcal{V}^{0}))$ for a suitable $t^{*}<1$ (we choose $t^{*}<1$ so that the involved constants do not depend on time). We determine the sequence $\{\mathcal{V}_{k}\}_{k=1}^{\infty}$ by solving the following set of linear hyperbolic PDEs
\begin{eqnarray}
\mathcal{L}_{k}\mathcal{V}_{k+1}=\mathcal{F}_{k},
\end{eqnarray}
where $\mathcal{L}_{k}:=\mathcal{L}[\mathcal{V}_{k}]$ and $\mathcal{F}_{k}=\mathcal{F}[\mathcal{V}_{k}]$. Notice $N_{k}=N[\mathcal{V}_{k}]$, $X_{k}=X[\mathcal{V}_{k}]$, and $\varphi_{k}=\varphi[\mathcal{V}_{k}]$ are determined by solving the associated elliptic equations (\ref{eq:elliptic},\ref{eq:elliptic2},\ref{eq:elliptic3}). Set $\mathcal{V}_{1}=\mathcal{V}^{0}_{1}$ and $\mathcal{F}_{1}=0$. Now we invoke the energy estimates of lemma (2) and apply them to the above set of linear hyperbolic PDEs. As long as $\{\mathcal{V}_{k^{'}}\}_{k^{'}=1}^{k}$ takes values in $B^{s}_{R}(\mathcal{V}^{0})$, the following energy estimate holds 
\begin{eqnarray}
\label{eq:energyapplied}
\sqrt{E_{s}(\mathcal{V}_{k+1};t)}\leq C_{R}\left(\sqrt{E_{s}(\mathcal{V}_{k+1};0)}+||\mathcal{F}_{k}||_{L^{1}([0,t];\mathcal{H}^{s}}\right)
\end{eqnarray}
for a $t<1$ and a constant $C_{R}$ dependent on $||\mathcal{V}_{k}||_{\mathcal{H}^{s}}$. Now, we have control on the initial sequence $\{\mathcal{V}^{0}_{k}\}_{k=1}^{\infty}$. In particular it is a Cauchy sequence in $B^{s}_{R}(\mathcal{H}^{s})$. In fact we restrict it within a smaller ball of radius $R/4$. More specifically 
\begin{eqnarray}
\mathcal{V}^{0}_{k}\in B^{s}_{R/4}(\mathcal{V}^{0}),~\forall k\geq 1,\\
C_{R}||\mathcal{V}^{0}_{k}-\mathcal{V}^{0}_{k^{'}}||_{\mathcal{H}^{s}}<\frac{R}{4}~\forall k,k^{'}\geq 1.
\end{eqnarray}
We need to prove that the sequence $\{\mathcal{V}_{k}\}_{k=1}^{\infty}$ converges in $\mathcal{C}\left([0,t^{*}];B^{s}_{R}(\mathcal{V}^{0})\right)$ and the limit solves the gauge fixed Einstein-Yang-Mills evolution equations for a suitable $t^{*}<1$. The major difficulty in the PDE setting is that the closed and bounded balls are not compact in infinite dimensions. We accomplish this goal by the following three steps.\\
\textbf{Lemma 5:} \textit{There exists a time $t^{*}\in (0,1)$ such that $\{\mathcal{V}_{k}\}_{k=1}^{\infty}\subset L^{\infty}([0,t^{*}];B^{s}_{R}(\mathcal{V}^{0}))$.}

\textbf{Proof:}
This is equivalent to proving that there exists a time $t^{*}\in (0,1)$ such that if $\{\mathcal{V}_{k^{'}}\}_{k^{'}=1}^{k}\subset L^{\infty}([0,t^{*}];B^{s}_{R}(\mathcal{V}^{0}))$, then $\mathcal{V}_{k+1}\in L^{\infty}([0,t^{*}];B^{s}_{R}(\mathcal{V}^{0}))$. For this purpose we use the following difference equation 
\begin{eqnarray}
\mathcal{L}_{k}(\mathcal{V}_{k+1}-\mathcal{V}_{1})=\mathcal{F}_{k}-\mathcal{L}_{k}\mathcal{V}_{1}
\end{eqnarray}
apply the energy estimate (\ref{eq:energyapplied}) to yield 
\begin{eqnarray}
||\mathcal{V}_{k+1}-\mathcal{V}_{1}||_{L^{\infty}([0,t^{*}];\mathcal{H}^{s})}\\\nonumber \leq C_{R}\left(||\mathcal{V}^{0}_{k+1}-\mathcal{V}^{0}_{1}||_{\mathcal{H}^{s}}+||\mathcal{F}_{k}||_{L^{1}([0,t^{*}];\mathcal{H}^{s})}+||\mathcal{L}_{k}\mathcal{V}_{1}||_{L^{1}([0,t^{*}];\mathcal{H}^{s})}\right)\\\nonumber 
=C_{R}\left(||\mathcal{V}^{0}_{k+1}-\mathcal{V}^{0}_{1}||_{\mathcal{H}^{s}}+||\mathcal{F}_{k}||_{L^{1}([0,t^{*}];\mathcal{H}^{s})}+||\mathcal{L}_{k}\mathcal{V}^{0}_{1}||_{L^{1}([0,t^{*}];\mathcal{H}^{s})}\right)
\end{eqnarray}
Now, by construction, we have $C_{R}||\mathcal{V}^{0}_{k+1}-\mathcal{V}^{0}_{1}||_{\mathcal{H}^{s}}<R/4$. Now we notice that $\int_{0}^{t^{*}}||\mathcal{F}_{k}||_{\mathcal{H}^{s}}dt\leq t^{*}||\mathcal{F}_{k}||_{L^{\infty}([0,t^{*}];\mathcal{H}^{s})}$ and $\int_{0}^{t^{*}}||\mathcal{L}_{k}\mathcal{V}^{0}_{1}||_{\mathcal{H}^{s}}dt\leq t^{*}||\mathcal{L}_{k}\mathcal{V}^{0}_{1}||_{L^{\infty}([0,t^{*}];\mathcal{H}^{s}})$. Therefore using the argument of lemma $3$ ($\mathcal{F}$ is a norm preserving map in a sense that $||\mathcal{F}(\mathcal{V}_{k})||_{\mathcal{H}^{s}}\lesssim ||\mathcal{V}_{k}||_{\mathcal{H}^{s}}$) there exists a time $t^{*}$ satisfying $0<t^{*}<1$ and such that the following holds 
\begin{eqnarray}
C_{R}t^{*}||\mathcal{F}_{k}||_{L^{\infty}([0,t^{*}];\mathcal{H}^{s})}<\frac{R}{4},~C_{R}t^{*}||\mathcal{L}_{k}\mathcal{V}^{0}_{1}||_{\mathcal{H}^{s}}<R/4
\end{eqnarray}
since $\{\mathcal{V}_{k^{'}}\}_{k^{'}=1}^{k}\subset L^{\infty}([0,t^{*}];B^{s}_{R}(\mathcal{V}^{0}))$. This yields 
\begin{eqnarray}
||\mathcal{V}_{k+1}-\mathcal{V}_{1}||_{L^{\infty}([0,t^{*}];\mathcal{H}^{s})}< 3R/4.
\end{eqnarray}
However by construction $\mathcal{V}_{1}=\mathcal{V}^{0}_{1}$ and therefore $\mathcal{V}_{1}\in B^{s}_{R/4}(\mathcal{V}^{0})$ . This yields $\mathcal{V}_{k+1}\in L^{\infty}([0,t^{*}];B^{s}_{R}(\mathcal{V}^{0}))$ for a suitable time $t^{*}\in (0,1)$.$~~~~~~~~~~~~~~~~~~~~~~~~~~~~~~~~~~~~~\square$ 

\textbf{Lemma 6:} \textit{There exists a time $t^{*}\in (0,1)$ such that $\{\mathcal{V}_{k}\}_{k=1}^{\infty}$} converges sub-sequentially in $L^{\infty}([0,t^{*}];\mathcal{H}^{1})$.

\textbf{Proof:} Now we construct the following equation 
\begin{eqnarray}
\mathcal{L}_{k}(\mathcal{V}_{k+1}-\mathcal{V}_{k^{'}+1})=\mathcal{F}_{k}-\mathcal{F}_{k^{'}}+(\mathcal{L}_{k^{'}}-\mathcal{L}_{k})\mathcal{V}_{k^{'}+1}
\end{eqnarray}
and apply the lowest order energy estimate (lemma 1) to yield
\begin{eqnarray}
||\mathcal{V}_{k+1}\nonumber-\mathcal{V}_{k^{'}+1}||_{L^{\infty}([0,t^{*}];\mathcal{H}^{1})}\\\nonumber 
\leq C_{R}\left(||\mathcal{V}^{0}_{k+1}-\mathcal{V}^{0}_{k^{'}+1}||_{\mathcal{H}^{1}}\nonumber+||\mathcal{F}_{k}-\mathcal{F}_{k^{'}}||_{L^{1}([0,t^{*}];\mathcal{H}^{1})}+||(\mathcal{L}_{k^{'}}-\mathcal{L}_{k})\mathcal{V}_{k^{'}+1}||_{L^{1}([0,t^{*}];\mathcal{H}^{1})}\right).
\end{eqnarray}
Now we use the Lipsitz property of the map $\mathcal{F}:\mathcal{H}^{1}\to \mathcal{H}^{1}$ from lemma 4 to yield 
\begin{eqnarray}
||\mathcal{V}_{k+1}-\mathcal{V}_{k^{'}+1}||_{L^{\infty}([0,t^{*}];\mathcal{H}^{1})}\\\nonumber 
\leq C_{R}\left(||\mathcal{V}^{0}_{k+1}-\mathcal{V}^{0}_{k^{'}+1}||_{\mathcal{H}^{1}}+t^{*}C_{L}||\mathcal{V}_{k}\nonumber-\mathcal{V}_{k^{'}}||_{L^{\infty}([0,t^{*}];\mathcal{H}^{1})}\right.\\\nonumber
\left.+t^{*}||(\mathcal{L}_{k^{'}}-\mathcal{L}_{k})\mathcal{V}_{k^{'}+1}||_{L^{\infty}([0,t^{*}];\mathcal{H}^{1})}\right).
\end{eqnarray}
Now, given $N_{k}=N(\mathcal{V}_{k}), X_{k}=X(\mathcal{V}_{k}), \varphi_{k}=\varphi(\mathcal{V}_{k})\in L^{\infty}([0,t^{*}];H^{s+1})$, using the product estimate (\ref{eq:product1}) with $t_{1}=1$, $t_{2}=s-2$ and $k=0$, we may estimate $||(\mathcal{L}_{k^{'}}-\mathcal{L}_{k})\mathcal{V}_{k^{'}+1}||_{L^{\infty}([0,t^{*}];\mathcal{H}^{1})}$ as 
\begin{eqnarray}
||(\mathcal{L}_{k^{'}}-\mathcal{L}_{k})\mathcal{V}_{k^{'}+1}||_{L^{\infty}([0,t^{*}];\mathcal{H}^{1})}\leq C^{'}_{L}||\mathcal{V}_{k}\nonumber-\mathcal{V}_{k^{'}}||_{L^{\infty}([0,t^{*}];\mathcal{H}^{1})}||\mathcal{V}_{k^{'}+1}||_{L^{\infty}([0,t^{*}];\mathcal{H}^{s})}.
\end{eqnarray}
This works out in a consistent manner since we have established in the previous lemma (lemma 5) that $\{\mathcal{V}_{k}\}_{k=1}^{\infty}\subset L^{\infty}([0,t^{*}];B^{s}_{R}(\mathcal{V}^{0}))$. Now we can choose a $t^{*}\in (0,1)$ (may decrease it if required) such that $t^{*}C_{L}C^{2}_{R}<\frac{1}{2}$. The previous inequality becomes 
\begin{eqnarray}
||\mathcal{V}_{k+1}-\mathcal{V}_{k^{'}+1}||_{L^{\infty}([0,t^{*}];\mathcal{H}^{1})}\\\nonumber 
\leq C_{R}||\mathcal{V}^{0}_{k+1}-\mathcal{V}^{0}_{k^{'}+1}||_{\mathcal{H}^{1}}+\frac{1}{2}||\mathcal{V}_{k}\nonumber-\mathcal{V}_{k^{'}}||_{L^{\infty}([0,t^{*}];\mathcal{H}^{1})}.
\end{eqnarray}
Setting $k^{'}=k-1$ we obtain 
\begin{eqnarray}
||\mathcal{V}_{k+1}-\mathcal{V}_{k}||_{L^{\infty}([0,t^{*}];\mathcal{H}^{1})}\\\nonumber 
\leq C_{R}||\mathcal{V}^{0}_{k+1}-\mathcal{V}^{0}_{k}||_{\mathcal{H}^{1}}+\frac{1}{2}||\mathcal{V}_{k}\nonumber-\mathcal{V}_{k-1}||_{L^{\infty}([0,t^{*}];\mathcal{H}^{1})}
\end{eqnarray}
which yields 
\begin{eqnarray}
\sum_{k=2}^{\infty}||\mathcal{V}_{k+1}-\mathcal{V}_{k}||_{L^{\infty}([0,t^{*}];\mathcal{H}^{1})}\\\nonumber
\leq 2C_{R}\sum_{k=1}^{\infty}||\mathcal{V}^{0}_{k+1}-\mathcal{V}^{0}_{k}||_{\mathcal{H}^{1}}\nonumber+||\mathcal{V}_{2}-\mathcal{V}_{1}||_{L^{\infty}([0,t^{*}];\mathcal{H}^{1})}.
\end{eqnarray}
Now we have control over the initial sequence (a Cauchy sequence) and therefore we can extract a sub-sequence to yield $2C_{R}\sum_{k=1}^{\infty}||\mathcal{V}^{0}_{k+1}-\mathcal{V}^{0}_{k}||_{\mathcal{H}^{1}}<\frac{R}{2}$ which leads to 
\begin{eqnarray}
\sum_{k=2}^{\infty}||\mathcal{V}_{k+1}-\mathcal{V}_{k}||_{L^{\infty}([0,t^{*}];\mathcal{H}^{1})}\leq R
\end{eqnarray}
concluding the sub-sequential convergence of $\{\mathcal{V}_{k}\}_{k=1}^{\infty}$ in $L^{\infty}([0,t^{*}];\mathcal{H}^{1})$. $~~~~~~~~~~~~~~~\square$

The remaining steps are showing that the limit is a solution of the gauge fixed Einstein-Yang-Mills field equations and that it lies in the space $\mathcal{C}([0,t^{*}];\mathcal{H}^{s}),~s>\frac{n}{2}+1$. To accomplish this we need to use the notion of weak continuity. The space of weakly continuous functions on $[0,t^{*}]$ with values in $H^{s}$ is defined as follows 
\begin{eqnarray}
C_{W}([0,t^{*}];H^{s}):=\left\{f\in H^{s}|\langle f,\Psi\rangle\in C([0,t^{*}])~\forall\Psi\in H^{-s}\right\},
\end{eqnarray}
Here $H^{-s}$ is the dual space of $H^{s}$. We invoke the following lemma from \cite{andersson2003elliptic} proof of which follows using a standard `$\epsilon/3$' argument.

\textbf{Lemma \cite{andersson2003elliptic}:}  \textit{Let $\{f_{k}\}_{k=1}^{\infty}\subset L^{\infty}([0,t^{*}];H^{s}),~s>0$ be a bounded sequence and assume $\{f_{k}\}_{k=1}^{\infty}$ is Cauchy in $L^{\infty}([0,t^{*}];H^{s^{'}})$ for some $s^{'}\in [0,s)$. Then $\exists f\in C_{W}([0,t^{*}];H^{s})\cap L^{\infty}([0,t^{*}];H^{s})$ such that $\{f_{k}\}_{k=1}^{\infty}$ converges weakly to $f$ uniformly in $L^{\infty}([0,t^{*}]$.}

Utilizing this lemma together with lemma 6, we obtain $\mathcal{V}_{k}\to \mathcal{V}\in C_{W}([0,t^{*}];H^{s})$. Now using this result, we show that this limit $\mathcal{V}$ satisfies the evolution equation $\mathcal{L}\mathcal{V}=\mathcal{F}$.

\textbf{Lemma 7:} \textit{Weak limit $\mathcal{V}$ of $\mathcal{V}_{k}$ solves $\mathcal{L}\mathcal{V}=\mathcal{F}$}

\textbf{Proof:} Let us write the following 
\begin{eqnarray}
\mathcal{L}[\mathcal{V}]\mathcal{V}-\mathcal{F}[\mathcal{V}]&=&(\mathcal{L}[\mathcal{V}]\mathcal{V}-\mathcal{L}[\mathcal{V}_{k}]\mathcal{V})\nonumber+\mathcal{L}[\mathcal{V}_{k}](\mathcal{V}-\mathcal{V}_{k+1})+(\mathcal{L}[\mathcal{V}_{k}]\mathcal{V}_{k+1}-\mathcal{F}[\mathcal{V}])\\\nonumber 
&=&(\mathcal{L}[\mathcal{V}]\mathcal{V}-\mathcal{L}[\mathcal{V}_{k}]\mathcal{V})\nonumber+\mathcal{L}[\mathcal{V}_{k}](\mathcal{V}-\mathcal{V}_{k+1})+(\mathcal{F}[\mathcal{V}_{k}]-\mathcal{F}[\mathcal{V}])
\end{eqnarray}
Now similar to the calculations associated with the previous lemma, we may estimate the first term as 
\begin{eqnarray}
||\mathcal{L}[\mathcal{V}]\mathcal{V}-\mathcal{L}[\mathcal{V}_{k}]\mathcal{V}||_{L^{\infty}([0,t^{*}];\mathcal{H}^{0})}\lesssim ||\mathcal{V}-\mathcal{V}_{k}||_{L^{\infty}([0,t^{*}];\mathcal{H}^{s})}||\mathcal{V}||_{L^{\infty}([0,t^{*}];\mathcal{H}^{1})}
\end{eqnarray}
but $\mathcal{V}_{k}$ converges to $\mathcal{V}$ in $L^{\infty}([0,t^{*}];\mathcal{H}^{s})$ as $k\to\infty$. Therefore $(\mathcal{L}[\mathcal{V}]\mathcal{V}-\mathcal{L}[\mathcal{V}_{k}]\mathcal{V})\to 0$ in $L^{\infty}([0,t^{*}];\mathcal{H}^{0})$ as $k\to\infty$. Similarly using Lipsitz property of $\mathcal{F}$, it follows that $(\mathcal{F}[\mathcal{V}_{k}]-\mathcal{F}[\mathcal{V}])\to 0$ in $L^{\infty}([0,t^{*}];\mathcal{H}^{1})$ as $k\to\infty$. There is a little subtlety in the term $\mathcal{L}[\mathcal{V}_{k}](\mathcal{V}-\mathcal{V}_{k+1})$. We want to show that this term approaches 0 in $L^{\infty}([0,t^{*}];\mathcal{H}^{0})$. In order for this to happen terms of the type $g^{ij}_{k}\partial_{i}\partial_{j}(g-g_{k+1})$ and $g^{ij}_{k}\partial_{i}\partial_{j}(A-A_{k+1})$ should converge in $H^{-1}$. But this is immediate using the product estimates (\ref{eq:product1},\ref{eq:product2}) and the duality of $H^{1}$ and $H^{-1}$ i.e., the multiplication $H^{s}\times H^{-1}\to H^{-1}$ is continuous for $s>\frac{n}{2}+1$. On the other hand, we notice that the left hand side is independent of $k$. Therefore, $\mathcal{L}[\mathcal{V}](\mathcal{V})=\mathcal{F}[\mathcal{V}]$. $~~~~~~~~~~~~~~~~~~~~~~~~~~~~~~~~~~~~~~~~~~~~~~~~~~~~~~~~~~~~~~~~~~~~~~~~~~\square$
\\
Now we prove the `in time' continuity of the solutions.

\textbf{Lemma 8:} \textit{Solution $\mathcal{V}$ is continuous as a map $t\mapsto \mathcal{V}(t)$.}

\textbf{Proof:} We have established that $t\mapsto \mathcal{V}(t)$ is weakly continuous. In order to prove strong continuity we need to prove left and right continuity at every point in the interval of local existence i.e., $[0,t^{*}]$. In other words, we must show that 
\begin{eqnarray}
\lim_{h\to0} \sup ||\mathcal{V}(t+h)-\mathcal{V}(t)||_{\mathcal{H}^{s}}=0=\lim_{h\to0}\sup ||\mathcal{V}(t-h)-\mathcal{V}(t)||_{\mathcal{H}^{s}}.
\end{eqnarray}
In this case, we need only show $\lim_{h\to 0}\sup||\mathcal{V}(t+h)-\mathcal{V}(t)||_{\mathcal{H}^{s}}=0$ since the argument can be repeated by reversing the orientation of time. By a re-parametrization of the time, we need only show the right continuity at $t=0$ i.e., $\lim_{t\to 0}\sup||\mathcal{V}(t)-\mathcal{V}(0)||_{\mathcal{H}^{s}}=0$. This may be done using a lemma of functional analysis (\cite{rudin1991functional}) which states that if $y_{n}\rightharpoonup y$ and $||y_{n}||\to ||y||$, then $||y_{n}-y||\to 0$. Now, we have shown earlier that $\mathcal{V}(t)\in C_{W}([0,t^{*}];\mathcal{H}^{s})$. Therefore we need only show that $\lim_{t\to0}\sup||\mathcal{V}(t)||_{\mathcal{H}^{s}}=||\mathcal{V}(0)||_{\mathcal{H}^{s}}$. But this holds by the energy estimates derived in lemma 1 and 2. Therefore, we obtain $\mathcal{V}\in \mathcal{C}([0,t^{*}];\mathcal{H}^{s}),~s>\frac{n}{2}+1$.$~~~~~~~~~~~~~~~~~~~~~~~~~~~~~~~~~~~~~~~~~~~~~~~~~~~~~~~~~~~~~~~~~~~~~~~~~\square$

\subsection{Uniqueness of a solution and its continuous dependence on the data} 
It is straightforward to establish the uniqueness of the solution $\mathcal{V}$. Let us assume that there exists two solutions $\mathcal{V}_{1}$ and $\mathcal{V}_{2}$ such that $\mathcal{V}_{1}(0)=\mathcal{V}_{2}(0)$. We need to show that $\mathcal{V}_{1}(t)=\mathcal{V}_{2}(t)~\forall t\in [0,t^{*}]$. But this follows by the energy inequalities and the calculations of the lemma 5 since $||\mathcal{V}_{1}(0)-\mathcal{V}_{2}(0)||_{\mathcal{H}^{s}}=0$. More precisely, one may obtain an energy inequality $\partial_{t}E_{12}\lesssim E_{12}$ for $\mathcal{V}_{1}(t)-\mathcal{V}_{2}(t)$ which yields $E_{12}(t)=0~\forall t\in [0,t^{*}]$ if $E_{12}(0)=0$ ($E_{12}(t)$ is the energy associated with the difference $\mathcal{V}_{1}(t)-\mathcal{V}_{2}(t)$).  For the continuous dependence of the unique solution on the initial data, we need to show that the map $\mathcal{V}(0)\mapsto \mathcal{V}(t;\mathcal{V}(0))$ is continuous. For this, we refer to \cite{andersson2003elliptic} instead of repeating it once again since it will be exactly same. This works out since the extra terms that arise due to the inclusion of the Yang-Mills fields preserve the crucial characteristics of the functional $\mathcal{F}$ e.g., $\mathcal{F}$ remains Lipschitz as shown in lemma 3 and 4. Instead, we focus on the elliptic estimates and conservation of gauges and constraints since now there are one additional constraint (Gauss law constraint for the electric field) and one additional gauge condition (generalized Coulomb gauge condition).   

\section{Elliptic Estimates}     
The remaining task is to show that the gauge variables $g,N,$ and $\varphi$ are determined in terms of $(g,k,A,\mathcal{E})$ and satisfy the regularity $H^{s+1}\times H^{s+1}\times H^{s+1}$. Let us recall the elliptic equations satisfied by these gauge variables. The lapse function $N$, the shift vector field $X$, and the shifted potential $\varphi$ satisfy
\begin{eqnarray}
-g^{ij}\nabla_{i}\nabla_{j}N\nonumber+\underbrace{\left(|k|^{2}_{g}+\frac{n-2}{n-1}g^{kl}\mathcal{E}^{a}~_{bk}\mathcal{E}^{b}~_{al}+\frac{1}{2(n-1)}g^{kl}g^{mn}F^{a}~_{bkm}F^{b}~_{aln}\right)}_{\geq 0}N\\\nonumber 
=1,
\end{eqnarray}
\begin{eqnarray}
\Delta_{g}X^{i}-R^{i}_{j}X^{j}+L_{X}V^{i}=-2k^{ij}\nabla_{j}N-2Ng^{ik}g^{lj}F^{a}~_{bkl} \mathcal{E}^{b}~_{aj}+\tau\nabla^{i}N\\\nonumber
+(2Nk^{jk}-2\nabla^{j}X^{k})(\Gamma[g]^{i}_{jk}-\Gamma[\gamma]^{i}_{jk}),
\end{eqnarray}
and 
\begin{eqnarray}
g^{ij}\nabla[\gamma]^{A}_{i}\nabla[\gamma]^{A}_{j}\varphi^{a}~_{b}+g^{ij}[\hat{A}_{i}-A_{i},\nabla^{A}_{j}\varphi]^{a}~_{b}+\nonumber
g^{ij}\nabla_{i}N\mathcal{E}^{a}~_{bj}-Ng^{ij}[A_{i},\mathcal{E}_{j}]^{a}~_{b}\\\nonumber+g^{ij}[\hat{A}_{i},N\mathcal{E}_{j}]^{a}~_{b}
+g^{ij}(\Gamma[g]^{k}_{ij}-\Gamma[\gamma]^{k}_{ij})N\mathcal{E}^{a}~_{bk}-X^{k}\nabla[\gamma]_{k}g^{ij}\nabla[\gamma]_{i}(A^{a}~_{bj}\\\nonumber-\hat{A}^{a}~_{bj})-X^{k}\nabla[\gamma]_{k}g^{ij}[\hat{A}_{i},A_{j}-\hat{A}_{j}]^{a}~_{b} +g^{ij}\nabla[\gamma]_{i}X^{k}\nabla[\gamma]_{k}A^{a}~_{bj}\\\nonumber+g^{ij}\nabla[\gamma]_{i}A^{a}~_{bk}\nabla[\gamma]_{j}X^{k}+g^{ij}A^{a}~_{bk}\nabla[\gamma]_{i}\nabla[\gamma]_{j}X^{k}-g^{ij}R[\gamma]^{l}_{jik}A^{a}~_{bl}X^{k} \\\nonumber+g^{ij}[\hat{A}_{i},A_{k}]^{a}~_{b}\nabla[\gamma]_{j}X^{k}+(2Nk^{ij}-\nabla^{i}X^{j}-\nabla^{j}X^{i})\nabla[\gamma]^{\hat{A}}_{i}(A^{a}~_{bj}-\hat{A}^{a}~_{bj})=0,
\end{eqnarray}
respectively. With $u=(N,X,\varphi)$, this system is of the following type 
\begin{eqnarray}
\mathcal{J}(g,k,A,Dg,DA)u=\mathcal{R}(g,k,A,Dg,DA),
\end{eqnarray}
where the second order elliptic operator $\mathcal{J}$ is of the form 
\begin{eqnarray}
\mathcal{J}(g,k,A,Dg,DA)u=g^{ij}\partial_{i}\partial_{j}u+b^{i}(g,k,A,Dg,DA)\partial_{i}u\nonumber+c(g,k,A,Dg,DA)u,
\end{eqnarray}
where $b$ and $c$ are smooth functionals of their arguments. If $\mathcal{J}$ is is an isomorphism between the function spaces $H^{s+1}$ and $H^{s-1}$, then the standard elliptic theory yields an estimate of the type 
\begin{eqnarray}
||u||_{H^{s+1}}\leq C(||\mathcal{J}u||_{H^{s-1}}+||u||_{H^{s-1}}).
\end{eqnarray}
Following the isomorphism property of the operator $\mathcal{J}$ and the compact embedding of $H^{s+1}$ into $H^{s-1}$ (for compact manifolds), one may obtain an estimate of the type (see \cite{andersson2003elliptic} for the logic behind this argument) 
\begin{eqnarray}
||u||_{H^{s+1}}\leq C||\mathcal{J}u||_{H^{s-1}}.
\end{eqnarray}
Counting the number of derivatives and using the  multiplication property of the Sobolev spaces, one controls the source term $||\mathcal{J}u||_{H^{s-1}}=||\mathcal{R}||_{H^{s-1}}$ yielding $u\in H^{s+1}$ i.e., $(N,X,\varphi)\in H^{s+1}\times H^{s+1}\times H^{s+1}$ for $s>\frac{n}{2}+1$.

The remaining task is to argue that the operator $\mathcal{J}$ is an isomorphism between the suitable function spaces. We consider each equation separately. First we show that the kernel of the map $\mathcal{Q}:N\mapsto -g^{ij}\nabla_{i}\nabla_{j}N\nonumber+\left(|k|^{2}_{g}+\frac{n-2}{n-1}g^{kl}\mathcal{E}^{a}~_{bk}\mathcal{E}^{b}~_{al}+\frac{1}{2(n-1)}g^{kl}g^{mn}F^{a}~_{bkm}F^{b}~_{aln}\right)N$ is trivial. A straightforward calculation by integration by parts yields $\ker(\mathcal{Q})=\{0\}$ since the entity under the bracket has good sign. The case of the shift vector field is slightly more subtle. Notice that $\Delta_{g}X^{i}-R^{i}_{j}X^{j}$ contains a second derivative term of the metric of the type $L_{X}V^{i}$ and therefore requires a $H^{s+2}$ regularity of $X$ for the metric to be in $H^{s}$. However, in spatial harmonic gauge (\ref{eq:sh}), this dangerous second derivative (of metric) term vanishes point-wise sense since $V^{i}\equiv 0$ in this gauge. For the detailed calculations, see lemma $3.2$ of \cite{andersson2003elliptic}. The isomorphism property of the map $X^{i}\mapsto \Delta_{g}X^{i}-R^{i}_{j}X^{j}+L_{X}V^{i}+2\nabla^{j}X^{k}(\Gamma[g]^{i}_{jk}-\Gamma[\gamma]^{i}_{jk})$ holds if one assumes $||g-\gamma||_{H^{s}}<\epsilon$ (this construction is possible since $C^{\infty}$ is dense in Sobolev spaces). This was proved in \cite{andersson2003elliptic} by considering the pull back of the tensor fields along the flow of the shift vector field and assuming $\gamma$ to have negative sectional curvature. Later \cite{moncrief2019could} sketched an argument (see their appendix) relaxing the restriction of $\gamma$ to have negative sectional curvature. Therefore, in the current context, the map $X^{i}\mapsto \Delta_{g}X^{i}-R^{i}_{j}X^{j}+L_{X}V^{i}+2\nabla^{j}X^{k}(\Gamma[g]^{i}_{jk}-\Gamma[\gamma]^{i}_{jk})$ is an isomophism between $H^{s+1}$ and $H^{s-1}$ for $||g-\gamma||_{H^{s}}<\epsilon$ for $\epsilon>0$.

A similar problem arises in the case of the elliptic equation for the Yang-Mills gauge variable $\varphi$. In order to obtain a unique $\varphi=\varphi(g,k,A,\mathcal{E})$,
one must establish the isomorphism property of the map $\varphi^{a}~_{b}\mapsto g^{ij}\nabla[\gamma]^{A}_{i}\nabla[\gamma]^{A}_{j}\varphi^{a}~_{b}+g^{ij}[\hat{A}_{i}-A_{i},\nabla^{A}_{j}\varphi]^{a}~_{b}$. If this map ceases to be an isomorphism, then the gauge orbit develops directions of tangency to the gauge slice or the gauge orbit intersects the slice more than once. This is turn introduces degeneracies and defies the whole purpose of gauge fixing.   This is the so called Gribov ambiguity for which a substantial amount of study exists in the literature \cite{moncrief1979gribov, gribov2001instability, chodos1980geometrical}. Exactly similar problems occur in the spatial harmonic gauge of gravity as well \cite{fischer1996quantum}. Similar to the gravity case if the background connection $\hat{A}$ is chosen close to the dynamical connection $A$ in $H^{s}$, then the map $\varphi^{a}~_{b}\mapsto g^{ij}\nabla[\gamma]^{A}_{i}\nabla[\gamma]^{A}_{j}\varphi^{a}~_{b}+g^{ij}[\hat{A}_{i}-A_{i},\nabla^{A}_{j}\varphi]^{a}~_{b}$ is proven to be an isomorphism (see \cite{singer1978some, moncrief1979gribov}). This can roughly be seen from the principal part of the equation. Notice that the operator $g^{ij}\nabla[\gamma]^{A}_{i}\nabla[\gamma]^{A}_{j}$ has a definite sign (negative) but the operator $g^{ij}[\hat{A}_{i}-A_{i},\nabla^{A}_{j}]$ does not. Therefore for small enough $(A-\hat{A})$ (in a suitable function space), the operator $g^{ij}\nabla[\gamma]^{A}_{i}\nabla[\gamma]^{A}_{j}+g^{ij}[\hat{A}_{i}-A_{i},\nabla^{A}_{j}]$ has an overall sign and consequently is invertible. Let us now rationalize the choice of $\hat{A}$ (or $\gamma$) such that $||A(t)-\hat{A}||_{H^{s}}<\epsilon$ (or $||g(t)-\gamma||_{H^{s}}<\epsilon$) for a sufficiently small $\epsilon$. Let us consider that the initial data $(g_{0},k_{0},A_{0},\mathcal{E}_{0})$ solves the constraint equations. Initially, the background connection and metric are chosen to be $A_{0}$ and $g_{0}$, respectively. This trivially satisfies the smallness conditions $||A(t)-\hat{A}||_{H^{s}}<\epsilon$ and $||g(t)-\gamma||_{H^{s}}<\epsilon$. Therefore, solving the elliptic equations, we also obtain the gauge variables $(N_{0},X_{0},\varphi_{0})$. Now we evolve the fields up to a time $\hat{t}$ within which the smallness condition is preserved i.e., the evolution time $\hat{t}$ is obtained by imposing the condition that $||g(\hat{t})-\gamma||_{H^{s}}\leq \epsilon/2$ and $||A(\hat{t})-\hat{A}||_{H^{s}}\leq \epsilon/2$. Once this range is exhausted, we choose the background connection and metric to be the instantaneous values of the dynamical connection and the metric at time $\hat{t}$ and repeat the evolution in the same way. The elliptic regularity of these equations yields unique $N,X,\varphi=N(g,k,A,\mathcal{E}), X(g,k,A,\mathcal{E}), \varphi(g,k,A,\mathcal{E})$ belonging to $H^{s+1}\times H^{s+1}\times H^{s+1}$.

\section{Conservation of constraints and gauges along the solution curve}
In addition to proving the local well-posedness of the Einstein-Yang-Mills evolution equations, we must show that the constraints and the gauge conditions are preserved along the solution curve $t\mapsto (g(t),k(t),A(t),\mathcal{E}(t),N(t),X(t),\varphi(t))$. In general, the equations  satisfied by the constraints (see \cite{rendall2008partial}) do not exhibit a particular type in an arbitrary gauge. This hinders one to prove the preservation of the constraints in a straightforward gauge-invariant way. However, one would expect that such preservation of the constraints must hold due to the Bianchi identities. Andersson and Moncrief \cite{andersson2003elliptic} showed that in CMCSH gauge, the constraints and the associated entities defining the gauge conditions (that vanish if the gauge conditions are imposed) satisfy a coupled nonlinear hyperbolic equation. Using standard energy arguments, they proved if the gauges and constraints are satisfied initially, then they are preserved along any solution curve. In the current case, we have additional terms arising from the Yang-Mills sector. The full set of gauges and constraints are defined as follows
\begin{eqnarray}
T:=\tr_{g}k-t,\\
S^{i}:=g^{kl}(\Gamma[g]^{i}_{kl}-\Gamma[\gamma]^{i}_{kl}),\\
H:=R(g)-|k|^{2}_{g}+(tr_{g}k)^{2}-\mathcal{E}\cdot \mathcal{E}-\frac{1}{2}F\cdot F-\nabla_{i}S^{i},\\
M_{i}:=\nabla_{i}tr_{g}k-2\nabla^{j}k_{ij}+2F_{ij}\cdot \mathcal{E}^{j},\\
G^{a}_{b}:=g^{ij}\nabla[\gamma]^{\hat{A}}_{j}(A^{a}~_{bi}-\hat{A}^{a}~_{bi}),\\
L^{a}~_{b}:=g^{ij}\nabla_{i}\mathcal{E}^{a}~_{bj}+g^{ij}[A_{i},\mathcal{E}_{j}]^{a}~_{b}
\end{eqnarray}
Explicit calculations using the equations of motion yield the following set of evolution equations for $(T,S^{i},H,M_{i},G,L)$
\begin{eqnarray}
\partial_{t}T=L_{X}T+NH,\\
\partial_{t}H=L_{X}H+\nabla_{i}NM^{i}+2NH\tr_{g}k+Nk^{ij}(\nabla_{i}S_{j}\nonumber+\nabla_{j}S_{i})+L_{S}(N\tr_{g}k)\\\nonumber -N\Delta_{g}T+2\nabla_{i}N\nabla^{i}T,\\
\partial_{t}S^{i}=L_{X}S^{i}+NM^{i},\\
\partial_{t}M_{i}=L_{X}M_{i}+H\nabla_{i}N+N\tr_{g}k M_{i}-N(\Delta_{g}S_{i}-\underbrace{R_{ij}S^{j}}_{I})\\\nonumber +N\tr_{g}k\nabla_{i}T+\nabla^{j}N(\nabla_{i}S_{j}+\nabla_{j}S_{i}),\\
\partial_{t}G^{a}~_{b}=X^{k}\nabla_{k}G^{a}~_{b}+NL^{a}~_{b},\\
\partial_{t}L^{a}~_{b}=X^{k}\nabla_{k}L^{a}~_{b}-N\Delta_{g}G^{a}~_{b}+N\tr_{g}k L^{a}~_{b}+\nabla^{i}N\nabla_{i}G^{a}~_{b}\\\nonumber +N[A_{i},\nabla^{i}G]^{a}~_{b}+[L,\varphi]^{a}~_{b}
\end{eqnarray}
Notice that this set of equations describe a coupled hyperbolic system of equations for the variables $(T,S^{i},H,M_{i},G,L)$. Notice that these equations are exactly similar to those derived by \cite{andersson2003elliptic}. The only difference is the appearance of two additional equations from the Yang-Mills sector. Remarkably, the evolution equations for the Yang-Mills gauge conditions and constraints are decoupled from the gravity sector. The evolution equations are of the standard hyperbolic form.     
We may use the standard energy argument to show that if $(T(0)=0,S^{i}(0)=0,H(0)=0,M_{i}(0)=0,G(0)=0,L(0)=0)$, then $(T(t),S^{i}(t),H(t),M_{i}(t),G(t),L(t))\equiv (0,0,0,0,0,0)$ along the solution curve $t\mapsto (g(t),k(t),A(t),\mathcal{E}(t),N(t),X(t),\varphi(t))$. Let us define a wave equation type of energy for the above system of evolution equations as follows 
\begin{eqnarray}
E_{cg}:=\\
\frac{1}{2}\int_{\Sigma}\left(T^{2}+|\nabla T|^{2}_{g}+H^{2}+G^{a}~_{b}G^{b}~_{a}\nonumber+g^{ij}g^{kl}\nabla_{i}G^{a}~_{bk}\nabla_{j}G^{b}~_{al}\right.\\\nonumber 
\left.+L^{a}~_{b} L^{b}~_{a}+|M|^{2}_{g}+|S|^{2}_{g}+|\nabla S|^{2}_{g}\right)\mu_{g},
\end{eqnarray}
where the dot indicates a positive definite inner product on the fibers of the gauge bundle. 
In the evaluation of the time derivative of the energy functional, there will be one borderline problematic term that requires careful control using a product estimate. Naturally, the principal terms cancel each other via integration by parts. The Lie derivative terms may be handled by the integration by parts argument as well (similar to the procedure used in the proof of lemma 1). Assuming $(g,N)\in H^{s}\times H^{s+1},~s>\frac{n}{2}+1$ and the estimate satisfied the by Ricci tensor 
\begin{eqnarray}
||Ric||_{H^{s-2}}\lesssim ||g||_{H^{s}}(1+||g||_{H^{s}}),
\end{eqnarray}
we may control the borderline problematic term $||NR_{ij}S^{i}M^{j}||_{L^{1}}$ as follows 
\begin{eqnarray}
||NR_{ij}S^{i}M^{j}||_{L^{1}}\lesssim ||N||_{L^{\infty}}||Ric~ \nonumber S||_{L^{2}}||M||_{L^{2}}\\\nonumber 
\lesssim ||N||_{L^{\infty}}|| Ric||_{H^{s-2}}||S||_{H^{1}}||M||_{L^{2}}\\\nonumber 
\lesssim ||N||_{L^{\infty}}||g||_{H^{s}}(1+||g||_{H^{s}}||S||_{H^{1}}||M||_{L^{2}}.
\end{eqnarray}
Here we have once again used the product estimate (\ref{eq:product1}). Using Holder's inequality and the fact that $(k,A,X,\varphi)\in H^{s-1}\times H^{s}\times H^{s+1}\times H^{s+1},~s>\frac{n}{2}+1$, we
obtain following energy inequality
\begin{eqnarray}
\partial_{t}E_{cg}\lesssim E_{cg}
\end{eqnarray}
which yields 
\begin{eqnarray}
E_{cg}(0)=0\Rightarrow E_{cg}(t)=0~\forall t\in [0,t^{*}].
\end{eqnarray}
This completes the proof of the conservation of gauges and constraints. Now we can state the main theorem as follows\\ 
\textbf{Local Well-posedness Theorem:} \textit{Let $(g_{0},k_{0},A_{0},\mathcal{E}_{0})\in H^{s}\times H^{s-1}\times H^{s}\times H^{s-1},~s>\frac{n}{2}+1$ be the initial data for the Cauchy problem of the Einstein-Yang-Mills evolution equations (\ref{eq:evolution1}-\ref{eq:evolution4}) in Constant Mean extrinsic Curvature Spatial Harmonic Generalized Coulomb (CMCSHGC) gauge satisfying the constraint equations (\ref{eq:HC}-\ref{eq:GLCT}). This CMCSHGC Cauchy problem is well posed in $\mathcal{C}\left([0,t^{*}];H^{s}\times H^{s-1}\times H^{s}\times H^{s-1}\right)$. In particular, there exists a time $t^{*}>0$ dependent on $||g_{0}||_{H^{s}},||k_{0}||_{H^{s-1}},||A_{0}||_{H^{s}}, ||\mathcal{E}_{0}||_{H^{s-1}}$ such that the solution map $(g_{0},k_{0},A_{0},\mathcal{E}_{0})\mapsto (g(t),k(t),A(t),\mathcal{E}(t),N(t),X(t),\varphi(t))$ is continuous as a map 
\begin{eqnarray}
H^{s}\times H^{s-1}\times H^{s}\times H^{s-1}\\
\to H^{s}\times H^{s-1}\nonumber\times H^{s}\times H^{s-1}\times H^{s+1}\times H^{s+1} \times H^{s+1}.  
\end{eqnarray}
Let $t^{*}$ be the maximal time of existence of a solution to the CMCSHGC Cauchy problem with data $(g_{0},k_{0},A_{0},\mathcal{E}_{0})$, then either $t^{*}=\infty$ or 
\begin{eqnarray}
\lim_{t\to t^{*}}\sup \max\left(C(\Lambda_{t}),||D\hat{g}(t)||_{L^{\infty}},||DA(t)||_{L^{\infty}},||\mathcal{E}(t)||_{L^{\infty}}\right)=\infty.
\end{eqnarray}
In addition the gauges and constraints are preserved along the solution curve $t\mapsto (g(t),k(t),A(t),\mathcal{E}(t),N(t),X(t),\varphi(t))$ if they are satisfied initially.}

\section{Concluding Remarks}
We have established a local well-posedness theorem for the coupled Einstein-Yang-Mills equations without any symmetry assumption utilizing techniques developed by \cite{andersson2003elliptic}. This is a routine procedure before studying the global properties of solutions. There are, however, a few subtleties present in the current context of the Yang-Mills fields coupled to gravity. Firstly, the divergence of the Yang-Mills curvature $\nabla^{j}F^{a}~_{bij}$ that appears in the equation of motion should be expressed in terms of the background metric $\gamma$ instead of the dynamical metric $g$ since the latter produces a Ricci curvature term in the evolution equation of the electric field obstructing the closure of the energy estimates. Secondly, it is necessary to choose the gauge variable of the Yang-Mills sector to be the shifted potential $\varphi:=A_{0}-A\cdot X$ instead of the usual $A_{0}$ (even in the temporal gauge one must impose $\varphi=0$ instead of $A_{0}=0$) in order to close the energy estimates. Lastly, the Gribov degeneracies associated with the spatial harmonic and generalized Coulomb gauge conditions needed to be addressed. The remaining arguments of proving a local well-posedness result are handled by the technique developed by \cite{andersson2003elliptic}. 

The continuation criteria of the coupled Einstein-Yang-Mills system is the boundedness of the entity $\mathcal{W}(t):=\max\left(C(\Lambda_{t}),||D\hat{g}(t)||_{L^{\infty}},||DA(t)||_{L^{\infty}},||\mathcal{E}(t)||_{L^{\infty}}\right)$ i.e., as long as $\mathcal{W}(t)$ remains finite, the required Sobolev norms remain finite allowing one to continue the solution in the future time direction. Notice that $\mathcal{W}$ contains all the derivatives of the spacetime metric $\hat{g}$ and the connection $A$. However, one may be able to relax this condition up to a certain extent. In the pure gravity case, \cite{klainerman2010breakdown} was able to obtain continuation criteria that did not require all derivatives of the metric $\hat{g}$ but rather only time derivative of the spatial metric and spatial derivatives of the lapse function and the shift vector field. In the pure Yang-Mills case on the Minkowski background, Eardley and Moncrief \cite{eardley1982global2} showed that the boundedness of a point-wise gauge-invariant norm of the Yang-Mills curvature was sufficient to continue the solutions. Later they utilized a light-cone estimate technique to show that the point-wise norm of the Yang-Mills curvature can not blow up in finite time and therefore obtained the global existence; in fact, a similar result persists on a globally hyperbolic background as well (work in progress with Prof. Moncrief).  Recently, \cite{puskar2021} proved the global existence of critically non-linear wave fields on background curved spacetimes which is similar in spirit with our current course of action, that is, first prove a local existence theorem and then obtain a point-wise bound for the wavefield using the light cone estimate technique thereby satisfying the continuation criteria. Unlike pure Yang-Mills or wavefields propagating on a Minkowski spacetime (or a background spacetime), gravity coupled to the Yang-Mills system is substantially more delicate. One can not in general prove a global existence theorem for the coupled system but rather improve the continuation criteria. Utilizing Moncrief's light cone estimate technique for the coupled system, we hope to obtain a point-wise bound on the spacetime curvature (or its trace-less part the Weyl curvature) and Yang-Mills curvature given a bound on a certain entity such as the strain tensor associated with a unit normal field to a CMC hypersurface. The wave equations for space-time curvature or the Yang-Mills curvature contain non-linearities that satisfy the so-called \textit{null} condition (a cancellation occurs between terms that cannot be controlled by the associated energies) and therefore such an improved continuation criterion is naturally expected. We note that such results for the vacuum Einstein equations \cite{klainerman2010breakdown} and Einstein coupled to Klein-Gordon and Maxwell equations \cite{shao2011breakdown} exist in the literature. These results were obtained by a method that is substantially different from Moncrief's light cone estimate. We hope that we will be able to apply the light cone estimate technique to obtain an improved continuation criterion of the coupled Einstein-Yang-Mills system where the current work marks the completion of the first step.       

\section{Acknowledgement}
P.M would like to thank Prof. Vincent Moncrief for numerous useful discussions related to this project and for his help improving the manuscript. P.M would also like to thank Prof. Yau for igniting the interest in coupled Einstein-Yang-Mills system. This work was supported by CMSA at Harvard University.

\author{Puskar Mondal}
\address{Center of Methematical Sciences and Applications, Department of Mathematics, Harvard University}
\ead{puskar\_mondal@fas.harvard.edu}

\end{document}